\documentclass[11pt,a4paper]{article}
\usepackage{etex}
\usepackage[utf8]{inputenc}
\usepackage[TS1,T1]{fontenc}
\usepackage{xspace}
\usepackage[a4paper,tmargin=3truecm,bmargin=3truecm,rmargin=2.2truecm,lmargin=2.2truecm]{geometry}
\usepackage{amssymb,amsmath,mathtools}
\usepackage[all]{xy}
\usepackage{array,booktabs}
\usepackage{graphicx}
\usepackage{pifont}
\usepackage{slashed}
\usepackage{enumitem}
\usepackage[english]{babel}
\usepackage{hyperref}
\usepackage{lmodern}

\usepackage{amsthm,thmtools}

\theoremstyle{definition}

\theoremstyle{remark}

\usepackage[square,numbers,sort,compress,semicolon,merge]{natbib}
\usepackage{tikz}
\usetikzlibrary{arrows,shapes,shadows,positioning,calc,backgrounds,fit}

\tikzstyle{conceptbox} = [%
					outer sep=3pt,
					line width=0.5pt,
					draw=black,
					rectangle,
					font=\mdseries,
					rounded corners=1mm,
					fill=none
				]
\tikzstyle{flechestructure} = [%
					-stealth',
					draw=black,
					fill=black,
					line width=1pt,
				]
				
\newcommand{\centeredlines}[1]{{\setlength{\tabcolsep}{0pt}\setlength{\arrayrulewidth}{0pt}%
\begin{tabular}{c}#1\end{tabular}}}

\hypersetup{
plainpages=false,
colorlinks=true,
linkcolor=black, 
anchorcolor=black, 
citecolor=black, 
urlcolor=black, 
menucolor=black, 
filecolor=black, 
bookmarksopen=true,
bookmarksnumbered=true}

\numberwithin{equation}{section}

\hypersetup{
pdfauthor={Jordan François, Serge Lazzarini, Thierry Masson},
pdftitle={Gauge field theories: various mathematical approaches},
pdfsubject={},
pdfcreator={pdflatex},
pdfproducer={pdflatex},
pdfkeywords={}
}

\setlist[itemize]{label=--, topsep=5pt, itemsep=2pt, parsep=0pt}
\setlist[enumerate]{topsep=5pt, itemsep=2pt, parsep=0pt}
\setlist[enumerate,1]{label=\arabic*., ref=\arabic*}
\setlist[enumerate,2]{label=\alph*., ref=\theenumi.\alph*}
\setlist[description]{font=\rmfamily\bfseries, topsep=5pt, itemsep=2pt, parsep=0pt,leftmargin=1em}

\DeclareFontFamily{U}{matha}{\hyphenchar\font45}
\DeclareFontShape{U}{matha}{m}{n}{
      <5> <6> <7> <8> <9> <10> gen * matha
      <10.95> matha10 <12> <14.4> <17.28> <20.74> <24.88> matha12
      }{}
\DeclareSymbolFont{matha}{U}{matha}{m}{n}
\DeclareFontSubstitution{U}{matha}{m}{n}

\DeclareFontFamily{U}{mathb}{\hyphenchar\font45}
\DeclareFontShape{U}{mathb}{m}{n}{
      <5> <6> <7> <8> <9> <10> gen * mathb
      <10.95> mathb10 <12> <14.4> <17.28> <20.74> <24.88> mathb12
      }{}
\DeclareSymbolFont{mathb}{U}{mathb}{m}{n}
\DeclareFontSubstitution{U}{mathb}{m}{n}

\DeclareMathSymbol{\varstar}       {2}{matha}{"0F}
\DeclareMathSymbol{\convolution}   {2}{mathb}{"0A}

\newcommand{\omi}[1]{\buildrel { \buildrel{#1}\over{\vee} } \over .}

\newcommand{\cpi}{\pi^\circ}
\newcommand{\Act}{\mathcal{S}}
\newcommand{\rke}{\tau}
\newcommand{\LC}{\text{LC}}

\newcommand{\circoverset}[1]{\mathring{#1}}

\newcommand{\nabladot}{\circoverset{\nabla}}	
\newcommand{\omegadot}{\circoverset{\omega}}	
\newcommand{\Rdot}{\circoverset{R}}	

\newcommand{\dvol}{\textnormal{dvol}}

\newcommand{\hnabla}{{\widehat{\nabla}}}
\newcommand{\homega}{\widehat{\omega}}	
\newcommand\hR{\widehat{R}}
\newcommand\hg{\widehat{g}}
\newcommand\hF{\widehat{F}}

\newcommand{\tR}{{\widetilde{\mathrm{R}}}}

\newcommand{\bOmega}{{\overline{\Omega}}}

\newcommand{\nc}{noncommutative\xspace}
\newcommand{\NC}{Noncommutative\xspace}
\newcommand{\ncg}{noncommutative geometry\xspace}
\newcommand{\NCG}{Noncommutative geometry\xspace}

\newcommand\ensvide{{\varnothing}} 
\newcommand\exter{{\textstyle\bigwedge}} 

\newcommand\Ad{{\text{\textup{Ad}}}} 
\newcommand\ad{{\text{\textup{ad}}}} 
\newcommand\inv{{\text{\textup{Inv}}}} 
\newcommand\lie{{\text{\textup{Lie}}}}
\newcommand\hor{{\text{\textup{Hor}}}} 
\newcommand\bas{{\text{\textup{Basic}}}} 
\newcommand\equ{{\text{\textup{equ}}}} 
\newcommand\Id{{\text{\textup{Id}}}} 
\newcommand\der{{\text{\textup{Der}}}} 
\newcommand\sign{{\text{\textup{sign}}}}

\newcommand\dd{\text{\textup{d}}} 
\newcommand\ds{\text{\textup{s}}} 
\newcommand\hd{\widehat{\dd}} 

\newcommand{\bbbone}{{\text{\usefont{U}{dsss}{m}{n}\char49}}}

\DeclarePairedDelimiter\abs{\lvert}{\rvert}

\DeclarePairedDelimiterX\scprod[2]{\langle}{\rangle}{#1,#2}

\newcommand{\invast}{\varstar}
\newcommand{\dualast}{\ast}
\newcommand{\grast}{\bullet}

\newcommand{\hodgeast}{{\mathord{\star}}}

\newcommand{\pbast}{\ast}

\newcommand\cdotaction{\mathord{\cdot}}

\DeclareMathOperator{\Int}{Int} 
\DeclareMathOperator{\Out}{Out} 
\DeclareMathOperator{\Inn}{Inn} 
\DeclareMathOperator{\Aut}{Aut} 
\DeclareMathOperator{\Diff}{Diff} 
\DeclareMathOperator{\End}{End} 
\DeclareMathOperator{\Hom}{\mathsf{Hom}} 
\DeclareMathOperator{\Ker}{Ker} 
\renewcommand{\ker}{\Ker} 
\DeclareMathOperator{\tr}{tr} 

\newcommand\varnotation[1]{{\mathcal{#1}}}
\newcommand\algnotation[1]{{\mathbf{#1}}}
\newcommand\grnotation[1]{{\mathsf{#1}}}
\newcommand\lienotation[1]{{\mathbf{\mathsf{#1}}}}
\newcommand\evnotation[1]{{\mathnormal{#1}}}
\newcommand\ehnotation[1]{{\mathcal{#1}}}
\newcommand\modnotation[1]{{\boldsymbol{#1}}}

\newcommand{\tla}{{\lienotation{TLA}}}

\newcommand\sfX{{\mathsf X}}

\newcommand\varA{{\varnotation{A}}}

\newcommand\varE{{\varnotation{E}}}

\newcommand\varL{{\varnotation{L}}}
\newcommand\varM{{\varnotation{M}}}

\newcommand\varP{{\varnotation{P}}}

\newcommand\varU{{\varnotation{U}}}

\newcommand\algA{{\algnotation{A}}}

\newcommand\algzero{{\grnotation{0}}}
\newcommand\algun{{\grnotation{1}}}

\newcommand\lieA{{\lienotation{A}}}
\newcommand\lieB{{\lienotation{B}}}

\newcommand\lieL{{\lienotation{L}}}

\newcommand\evE{{\evnotation{E}}}

\newcommand\ehH{\ehnotation{H}}

\newcommand\modM{{\modnotation{M}}}

\newcommand\ka{{\mathfrak a}}
\newcommand\kb{{\mathfrak b}}
\newcommand\kg{{\mathfrak g}}
\newcommand\kh{{\mathfrak h}}

\newcommand\kp{{\mathfrak p}}

\newcommand\kD{{\mathfrak D}}
\newcommand\kS{{\mathfrak S}}

\newcommand\kX{{\mathfrak X}}
\newcommand\kY{{\mathfrak Y}}

\newcommand\ksl{{\mathfrak{sl}}}

\newcommand\ksu{{\mathfrak{su}}}

\newcommand\gC{{\mathbb C}}

\newcommand\gF{{\mathbb F}}

\newcommand\gH{{\mathbb H}}

\newcommand\gN{{\mathbb N}}

\newcommand\gR{{\mathbb R}}

\newcommand\caA{{\mathcal A}}
\newcommand\caB{{\mathcal B}}

\newcommand\caD{{\mathcal D}}

\newcommand\caG{{\mathcal G}}
\newcommand\caH{{\mathcal H}}

\newcommand\caL{{\mathcal L}}

\newcommand\caZ{{\mathcal Z}}

\begin{document}

\title{Gauge field theories: various mathematical approaches}
\author{Jordan François, Serge Lazzarini and Thierry Masson}
\date{}
\maketitle

\begin{center}
Aix Marseille Université, Université de Toulon, CNRS,\\
CPT, UMR 7332, Case 907, 13288 Marseille, France.
\end{center}

\bigskip
\begin{center}
To be published in the book\\
{\itshape Mathematical Structures of the Universe}\\ 
(Copernicus Center Press, Kraków, Poland, 2014)
\end{center}

\bigskip
\begin{abstract}
This paper presents relevant modern mathematical formulations for (classical) gauge field theories, namely, ordinary differential geometry,  \nc geometry, and transitive Lie algebroids. They provide rigorous frameworks to describe Yang-Mills-Higgs theories or gravitation theories, and each of them improves the paradigm of gauge field theories. A brief comparison between them is carried out, essentially due to the various notions of connection. However they reveal a compelling common mathematical pattern on which the paper concludes.
\end{abstract}

\newpage

\tableofcontents

\section{Introduction}

Since its inception in 1918, 1927 and 1929, through the pioneering work of Weyl, London, and Fock on electromagnetism, the idea of local symmetries, or gauge symmetries, has proven to be a decisive insight in the structure of fundamental interactions (for a general account see \textsl{e.g.} \cite{ORai97a} and references therein). The elaboration of these theories provides a spectacular example of convergence between physics and mathematics. In the early 1950's, while Yang and Mills proposed their idea of non abelian gauge fields (generalization of electromagnetism), Ehresmann developed the notion of connections on principal fiber bundles, which turns out to be the natural mathematical framework for Yang-Mills field theories.

In the 1960's, the elaboration of the Standard Model (SM) of particle physics has shown that three of the four fundamental interactions (electromagnetism, weak and strong interactions) can be modeled as abelian and Yang-Mills gauge fields, supplemented by a $\gC^2$-valued scalar field which generates masses for the gauge bosons through the spontaneous symmetry breaking mechanism (SSBM).
The discovery of the massive vector bosons $Z_\mu$, $W_\mu^\pm$ in 1983, and of the massive Higgs boson in 2012, confirms the relevance of the SM in its present formulation, in terms of the mathematics of connections. In this formulation, one has the correspondence between physical objects and mathematical structures given in Table~\ref{table-SM}. However it remains a weakness in this mathematical scheme. Indeed, the $\gC^2$-valued scalar field involved in the SSBM is, at the same time, a section of a (suitable) associated vector bundle~\cite{Trau79a, West80a, Ster94a}, and a boson, so that it is an ``hybrid structure'' belonging to the two rows of the table. Moreover, in this scheme, its scalar potential does not emerge from a natural mathematical construction.

The theory for the fourth fundamental interaction, gravitation, has been elaborated by Einstein within the framework of (pseudo-)riemannian geometry, and not as a gauge field theory. Later, this theory has been reformulated using connections on the frame bundle of space-time.  These reformulations have a richer structure (originating in the notion of soldering form) than bare Yang-Mills theories based on Ehresmann connections.

\begin{table}[t]
{\centering
\begin{tabular}{ccccc}
\toprule
\textbf{Kind of Particle} & & \textbf{Statistics} & & \textbf{Mathematical Structure} \\
\midrule
interaction particle &\ding{214}& 
boson
&\ding{214}& connection on principal fiber bundle\\
matter particle &\ding{214}&
 fermion 
 &\ding{214}& 
section of associated vector bundle
\\
\bottomrule
\end{tabular}
\par}
\caption{Correspondence between physical objects and mathematical structures in the usual formulation of the SM (see~\ref{sec Ehresmann connections}).}
\label{table-SM}
\end{table}

\smallskip
In this paper, we review and compare three mathematical frameworks suited to formulate gauge field theories, namely, ordinary differential geometry,  \nc geometry, and  the framework of transitive Lie algebroids. 

\NC geometry has been the first attempt to develop gauge field theories beyond the usual geometry of fiber bundles and connections. One of its first successes has been to propose gauge field theories in which scalar fields are part of the generalized notion of connection, and in which a naturally constructed Lagrangian produces a quadratic potential for these (new) fields, providing a SSBM in these models \cite{DuboKernMado90a, DuboKernMado90b, ConnLott90a}. Despite the success of the (re)construction of a \nc version of the SM \cite{ChamConnMarc07a}, \nc geometry has never been widely adopted as a new framework to model physics beyond the SM. The most important reason for the rejection of \nc geometry is certainly due to the mathematical skill required to master this new conceptual framework.

On the other side, the newly proposed framework of gauge field theories on transitive Lie algebroids \cite{FourLazzMass13a, LazzMass12a}, while giving rise to Yang-Mills-Higgs theories following the same successful recipes discovered in \nc geometry, is close enough to ordinary differential geometry and to usual algebraic structures, to permit to a wider audience to master this scheme. Moreover, contrary to \nc geometry in which the gauge group is the group of automorphisms of an associative algebra (this disqualifies $U(1)$-gauge field theories in \nc geometry), any gauge group of a principal fiber bundle can be used, since it is possible to consider the transitive Atiyah Lie algebroid associated to this principal fiber bundle.

An important point of this review paper is to compare, in both schemes, how scalar fields supplements naturally Yang-Mills fields in the corresponding notions of ``generalized'' connections. Then these fields belong to the first row of (a generalized version of) Table~\ref{table-SM}, as bosons and as part of  (generalized) connections. These schemes also provide a SSBM without requiring some extra (God given) inputs in the model: the associated scalar potential is given by the Lagrangian describing the dynamics of the fields of these generalized connections.

Another main point on which this paper focuses is the modeling of gravitation theories in terms of gauge field theories. A new recent way of thinking about symmetry reduction \cite{FourFranLazzMass14a} permits to clearly understand how the geometrical objects of the Einstein's theory of gravity are reconstructed after the decoupling of a gauge symmetry modeled in terms of Cartan connections. 

\smallskip
Let us emphasize an essential characterization of gauge field theories, as they are described and considered in this review paper.  Any theory written as the integral of a Lagrangian globally defined on a (space-time) manifold $\varM$ is necessarily invariant under diffeomorphisms. This is due to the fact that this Lagrangian must be invariant under any change of local coordinates on $\varM$ (at the price to introduce a non Minkowskian metric if necessary), and a diffeomorphism on $\varM$ is locally equivalent to a change of coordinates. This ``basic'' symmetry, in the sense that it is always required and also in the mathematical sense that it is governed by the ``base'' manifold $\varM$, is consistent with the terminology ``natural geometry'' put forward in \cite{KolaMichSlov93a}. In addition to this basic symmetry, the Lagrangian can be symmetric under more general transformations. Among them are the gauge symmetries, which, in the usual point of view, require an extra (non basic) structure. This extra structure is a principal fiber bundle $\varP$ in ordinary geometry: it is defined on top of $\varM$, but it can not be reconstructed from $\varM$ only. This is the essence of gauge symmetries, which are constrains on modeling of physical systems supplementing the basic space-time constrains.

\medskip
Gauge field theories are based on physical ideas which require essential mathematical structures in order to be elaborated. These basic ingredients can be listed as follows:
\begin{enumerate}
\item A space of local symmetries, (local in the sense that they depend on points in space-time): for instance this space is usually given by a so-called \emph{gauge group} (finite gauge transformations) or a \emph{Lie algebra} (infinitesimal gauge transformations).

\item An implementation of the symmetry on matter fields: it takes the form of a \emph{representation theory} associated to the natural mathematical structures of the theory.

\item A notion of derivation: that is the \emph{differential structure} on which equations of motion are written.

\item A replacement of ordinary derivations: this is the \emph{covariant derivative}, which encodes the physical idea of  ``minimal coupling'' between matter fields and gauge fields.

\item A way to write a gauge invariant Lagrangian density (up to a divergence term): this is the \emph{action functional}, from which the equations of motion are deduced.
\end{enumerate}

Let us emphasize that the three mathematical frameworks under consideration in this paper fulfill all these main features.
In order to get right away a direct comparison between the three, let us gather the listed items just down below.

\smallskip
In ordinary differential geometry which will be recalled in Section~\ref{sec-Ordinary differential geometry}, the fundamental mathematical structure is that of a $G$-principal fiber bundle $\varP$ over a smooth $m$-dimensional manifold $\varM$ which is usually expressed as the sequence $\xymatrix@1@C=15pt{{G} \ar[r] & {\varP} \ar[r]^-{\pi} & {\varM}}$. Then the ingredients are:
\begin{description}
\item[The gauge group:] this is the group of vertical automorphisms of $\varP$, denoted by $\caG(\varP)$.

\item[The representation theory:] any (linear) representation $\ell$ of $G$ on a vector space $\evE$ defines an associated vector bundle $\varE = \varP \times_\ell \evE$, and there is a natural action of $\caG(\varP)$ on the space of sections of $\varE$.

\item[The differential structure:] it consists into the (ordinary) de~Rham differential calculus.

\item[The covariant derivative:] any (Ehresmann) connection $1$-form $\omega$ on $\varP$ induces a covariant derivative $\nabla$ on sections of any associated vector bundles.

\item[The action functional:] in order to define an action functional, one needs an integration on the base manifold $\varM$, a Killing form on the Lie algebra $\kg$ of $G$, and the Hodge star operator associated to a metric on $\varM$. Then the action functional is written using the curvature of $\omega$.

\end{description}
An important aspect of this framework is that the connection, which contains the Yang-Mills gauge fields, is defined on the main structure (the principal fiber bundle $\varP$), and these fields couple to matter fields only when a representation is given.  In the same way, gauge transformations are defined on $\varP$, and they act on any object naturally introduced in the theory.

We will see in \ref{sec Cartan geometry} that Cartan connections can also be used, in replacement of Ehresmann connections, in particular to model gravitation as a gauge field theory.

\smallskip
In \ncg dealt with in Section~\ref{sec-NC}, the basic ingredient is an associative algebra $\algA$. Think of it as a replacement for the (commutative) algebra $C^\infty(\varM)$. Then one has:
\begin{description}
\item[The representation theory:] it consists to a right module $\modM$ over $\algA$. It is often required to be a projective finitely generated right module such that the theory is not empty.

\item[The gauge group:] this is the group $\Aut(\modM)$ of automorphisms of the right module. Contrary to ordinary differential geometry, it does depend on the representation space.

\item[The differential structure:] any differential calculus defined on top of $\algA$ can be used. There is no canonical construction here, and one has to make an explicit choice at this point. At least two important directions can be followed: consider the derivation-based differential calculus canonically associated to the algebra $\algA$ (see \ref{sec Derivation-based NCG}), or introduce supplementary structure to constitute a spectral triple $(\algA, \ehH, \caD)$ (see \ref{sec Spectral triples}).

\item[The covariant derivative:] it is a \nc connection, which is defined on $\modM$ with the help of the chosen differential calculus. In many situations, as in ordinary differential geometry, this covariant derivative can be equivalently described by a $1$-form in the chosen space of forms.

\item[The action functional:] it heavily depends on the choice of the differential calculus. For instance, using a derivation-based differential calculus, one can use some \nc counterparts of integration and Hodge star operator to construct a gauge invariant action based on the curvature of the connection. When a spectral triple is given, it is convenient to consider the spectral action associated to the Dirac operator $\caD$, which requires the Dixmier trace as a substitute for the integration.
\end{description}
Here, the connection (the generalized Yang-Mills fields) and the gauge transformations are defined \emph{acting on} matter fields, not at the level of the primary object $\algA$. This implies that the construction of a gauge field theory must take into account, \emph{at the very beginning}, the matter content. This way of thinking departs from the one in ordinary differential geometry, where gauge theories without matter fields can be considered. A way out is to particularize the right module as the algebra itself.

\smallskip
In the framework of transitive Lie algebroids to which Section~\ref{sec Transitive Lie algebroids} is devoted, the basic structure is a short exact sequence of Lie algebras and $C^\infty(\varM)$-modules, $\xymatrix@1@C=15pt{{\algzero} \ar[r] & {\lieL} \ar[r]^-{\iota} & {\lieA} \ar[r]^-{\rho} & {\Gamma(T\varM)} \ar[r] & {\algzero}}$, where $\rho$ satisfies some axioms (see \ref{sec Generalities on transitive Lie algebroids}). Think of it as an infinitesimal version of a principal fiber bundle $\xymatrix@1@C=15pt{{G} \ar[r] & {\varP} \ar[r]^-{\pi} & {\varM}}$. Then one has:
\begin{description}
\item[The differential structure:] it consists on a space of ``forms'' defined as multilinear antisymmetric maps from $\lieA$ to $\lieL$, equipped with a differential which takes into account the Lie structure on $\lieA$.

\item[The representation theory:] to any vector bundle $\varE$ over $\varM$, one can associate its transitive Lie algebroid of derivations, denoted by $\kD(\varE)$ (first order differential operators on $\varE$ whose symbol is the identity). Then a representation of $\lieA$ is a morphism of Lie algebroids $\lieA \to \kD(\varE)$.

\item[The gauge group:] given a representation as above, the gauge group is the group $\Aut(\varE)$ of vertical automorphisms of $\varE$. This depends on the vector bundle $\varE$. But \emph{infinitesimal} gauge transformations can be defined as elements of $\lieL$ (see~\ref{sec Gauge field theories}), independently on any representation of $\lieA$.

\item[The covariant derivatives:] there is a good notion of ``generalized connections'', which are defined as $1$-forms $\homega : \lieA \to \lieL$. Then a representation on $\varE$ induces an element in $\kD(\varE)$ associated to $\omega$. This is the covariant derivative.

\item[The action functional:] one can write a gauge invariant action functional using natural objects on $\lieA$, which consists into a metric, its Hodge star operator, a notion of integration along $\lieL$, and an integration on $\varM$.
\end{description}
As in \nc geometry, finite gauge transformations are only defined once a representation is given. But, as in ordinary differential geometry, a connection is intrinsically associated to the main structure (the short exact sequence), as well as are infinitesimal gauge transformations. Moreover, finite gauge transformations can be defined on any Atiyah Lie algebroid (see \ref{sec Gauge field theories}), which are  the natural transitive Lie algebroids to consider to get the closer generalizations of Yang-Mills field theories in this framework.

\smallskip
Althrough the three frameworks look quite different,  it will be shown that they present similarities from which a general scheme emerges. The latter can be summarized under the form of the sequence \eqref{eq-universal-sequence}, which ought to provide a general setting for treating gauge field theories at the classical level.

\section{Ordinary differential geometry}
\label{sec-Ordinary differential geometry}

Many textbooks explain in details the theory of fiber bundles and connections (see for instance \cite{Bert96a, KobaNomi96c, Naka90a,GockSchu89a}). We will suppose that the reader is quite familiar with these notions. Here, we will concentrate on ordinary (Ehresmann) connections on principal fiber bundles, a notion that will be generalized in the next two sections, and on the geometry of Cartan connections, which permits to consider Einstein theory of gravitation as a gauge theory.

\subsection{Ehresmann connections and Yang-Mills theory}
\label{sec Ehresmann connections}

Let 
\begin{equation}
\label{eq-principal-bundle}
\xymatrix@1{ {G} \ar[r] & {\varP} \ar[r]^-{\pi} & {\varM}}
\end{equation}
be a $G$-principal fiber bundle for a Lie group $G$ and a $m$-dimensional smooth manifold $\varM$. Denote by $\tR$ the right action of $G$ on $\varP$: $\tR_g(p) = p \cdotaction g$ for any $g \in G$ and $p \in \varP$. Let $\kg$ be the Lie algebra of $G$. A connection on $\varP$ can be characterized following two points of view.

The geometer says that a connection is a $G$-equivariant horizontal distribution $H\varP$ in the tangent bundle $T\varP$: for any $p \in \varP$, one supposes given a linear subspace $H_p\varP \subset T_p\varP$ such that $H_{p \cdotaction g}\varP = T_p \tR_g (H_p\varP)$. The curvature of the connection measures the failure for the distribution $H\varP$ to be integrable.

However, a dual equivalent algebraic setting is better suited to field theory. The distribution $H\varP$ is thus defined as the kernel of a $1$-form $\omega$ on $\varP$ with values in $\kg$. The algebraist then says that a connection on $\varP$ is an element $\omega \in \Omega^1(\varP) \otimes \kg$ such that
\begin{align}
\label{eq-relations-def-omega}
\omega(\xi^\varP) &= \xi, \text{\quad $\forall \xi \in \kg$},
&
{\tR}_g^\pbast \omega &= \Ad_{g^{-1}}\omega, \text{\quad $\forall g \in G$.}
\end{align}
where $\xi^\varP$ is the fundamental vector field on $\varP$ associated to the action $\tR_{e^{t\xi}}$. The curvature of $\omega$ is defined as the $2$-form $\Omega \in \Omega^2(\varP) \otimes \kg$ given by the Cartan structure equation $\Omega = \dd \omega + \tfrac{1}{2}[\omega, \omega]$ (where the graded bracket uses the Lie bracket in $\kg$). The space of connections is an affine space.

The gauge group $\caG(\varP)$ is the group of vertical automorphisms of $\varP$, which are diffeomorphisms $\Xi : \varP \to \varP$ which respect fibers and such that $\Xi(p \cdotaction g) = \Xi(p) \cdotaction g$ for any $p \in \varP$ and $g \in G$. This group acts by pull-back on forms, and it induces an action on the space of connections, \textsl{i.e.} $\Xi^\pbast \omega$ satisfies also \eqref{eq-relations-def-omega}. The gauge group can also be described as sections of the associated fiber bundle $\varP \times_\alpha G$ for the action $\alpha_g(h) = g h g^{-1}$ of $G$ on itself, and also as covariant maps $\Upsilon : \varP \to G$ satisfying $\Upsilon(p\cdotaction g) = g^{-1} \Upsilon(p) g$. Then a direct computation shows that (with $\dd$ the de~Rham differential on $\varP$)
\begin{align}
\label{eq-action-gauge-group-omega-Omega}
\Xi^\pbast \omega &= \Upsilon^{-1} \omega \Upsilon + \Upsilon^{-1} \dd \Upsilon,
&
\Xi^\pbast \Omega &= \Upsilon^{-1} \Omega \Upsilon.
\end{align}
The Lie algebra of infinitesimal gauge transformations is the space of sections of the associated vector bundle in Lie algebras $\Ad\varP = \varP \times_\Ad \kg$ for the $\Ad$ representation of $G$ on $\kg$.

The theory of fiber bundles tells us that $\varP$ can be locally trivialized by using a couple $(\varU, \phi)$, where $\varU \subset \varM$ is an open subset and $\phi : \varU \times G \to \pi^{-1}(\varU)$ is a isomorphism such that $\phi(x, gh) = \phi(x, g) \cdotaction h$ for any $x \in \varU$ and $g,h \in G$. Then $s : \varU \to \pi^{-1}(\varU)$ defined by $s(x) = \phi(x,e)$ is a local trivializing section, and one defines the local trivializations of $\omega$ and $\Omega$ on $\varU$ as
\begin{align}
\label{eq-local-description-A-F}
A&= s^\pbast \omega \in \Omega^1(\varU) \otimes \kg,
&
F&= s^\pbast \Omega \in \Omega^2(\varU) \otimes \kg.
\end{align}
As section of $\varP \times_\alpha G$, an element of the gauge group, can be trivialized into a map $\gamma : \varU \to G$, and its action $A \mapsto A^\gamma$ and $F \mapsto F^\gamma$ is given by the local versions of \eqref{eq-action-gauge-group-omega-Omega}:
\begin{align}
\label{eq-action-gauge-group-A-F}
A^\gamma &= \gamma^{-1} A \gamma + \gamma^{-1} \dd \gamma,
&
F^\gamma &= \gamma^{-1} F \gamma.
\end{align}

Let $\{ (\varU_i, \phi_i) \}_{i \in I}$ be a family of trivializations of $\varP$ such that $\bigcup_{i \in I} \varU_i = \varM$. On any $\varU_i \cap \varU_j \neq \ensvide$, there is then a map $g_{ij} = \varU_i \cap \varU_j  \to G$ such that $\phi_i(x, g_{ij}(x)) = \phi_j(x, e)$. Let us define the family $A_i = s_i^\pbast \omega$ and $F_i = s_i^\pbast \Omega$ for any $i \in I$. These forms satisfy the gluing relations
\begin{align}
\label{eq-gluing-relations-Ai-Fi}
A_j &=  g^{-1}_{ij} A_i g_{ij} + g^{-1}_{ij} \dd g_{ij},
&
F_j &=  g^{-1}_{ij} F_i g_{ij}.
\end{align}
These local expressions are those used in field theory, namely $A$ is the gauge potential and $F$ the field strength.

Note the similarity between active gauge transformations \eqref{eq-action-gauge-group-A-F} and gluing relations \eqref{eq-gluing-relations-Ai-Fi} (which are called ``passive gauge transformations''). A Lagrangian written in terms of $F$ and $A$ which is invariant under active gauge transformations is automatically compatible with the gluing relations, so that it is well defined everywhere on the base manifold.

The $2$-form $\Omega$ is horizontal, in the sense that it vanishes on vertical vector fields, and its is $(\tR, \Ad)$-equivariant, $\tR_g^\pbast \Omega = \Ad_{g^{-1}} \Omega$ for any $g \in G$. We say that $\Omega$ is tensorial of type $(\tR, \Ad)$. Such a form defines a form $\gF \in \Omega^2(\varM, \Ad\varP)$. The existence of $\gF$ can also be deduced from the homogeneous gluing relations \eqref{eq-gluing-relations-Ai-Fi} for the $F_i$'s. The $1$-form $\omega$ is not tensorial (the gluing relations of the $A_i$'s are inhomogeneous) so that it does does define a global form on $\varM$ with values in an associated vector bundle.

What we end up with is an equivalent description of these algebraic structures at three levels:
\begin{description}
\item[Globally on $\varP$:] $\omega \in \Omega^1(\varP)\otimes \kg$ is the connection $1$-form on $\varP$, which satisfies \eqref{eq-relations-def-omega} (equivariance and a vertical normalization), and $\Omega \in \Omega^2(\varP)\otimes \kg$ its curvature, which is tensorial of type $(\tR, \Ad)$. This is in general the preferred description for mathematicians. 

\item[Locally on $\varM$:] on any local trivialization $(\varU_i, \phi_i)$ of $\varP$, with associated local section $s_i(x) = \phi_i(x,e)$, the pull-back by $s_i$ defines the local descriptions $A_i$ and $F_i$ as in \eqref{eq-local-description-A-F}. These local descriptions are related from one trivialization to another by the gluing relations \eqref{eq-gluing-relations-Ai-Fi}. This is the preferred description for physicists, who define field theories in term of maps on space-time (the manifold $\varM$) to write down local Lagrangian.

\item[Globally on $\varM$:] The curvature is also a $2$-form $\gF$ globally defined on $\varM$, with values in an associated vector bundle. This description is not complete: the connection does not define a global $1$-form on $\varM$. Nevertheless, notice that the difference of two connections belongs to $\Omega^1(\varM, \Ad\varP)$. While incomplete, this description is the one that will be generalized in \ref{sec Basic structures in NCG} and \ref{sec Gauge field theories}. As we will see then, if one accepts to depart from ordinary differential geometry, a convenient space can be defined to consider a ``$1$-form'' to represent the connection $\omega$ in this description.
\end{description}

The connection $\omega$ defined on $\varP$ induces a ``connection'' on any associated vector bundle $\varE = \varP \times_\ell \evE$. From a geometric point of view, such a connection is a notion of parallel transport in the fibers along paths on the base manifold. Looking at the infinitesimal version of this parallel transport, one can define a derivation on the space $\Gamma(\varE)$ of smooth sections of $\varE$: this is the covariant derivative. In physics, matter fields are represented as elements in $\Gamma(\varE)$. Notice that the gauge group acts naturally on sections of any associated vector bundle, so that the symmetry is automatically implemented on any space of matter fields. Contrary to what will be described in \ref{sec Basic structures in NCG}, the gauge group is independent of the space $\Gamma(\varE)$.

Recall that in a vector bundle $\varE$, there is no canonical way to define a derivation of $\psi \in \Gamma(\varE)$ along a vector field $X \in \Gamma(T\varM)$. The connection is precisely the structure needed to define this ``derivation along $X$''. The covariant derivative defined by $\omega$ associates to any $X \in \Gamma(T\varM)$ a linear map $\nabla_X : \Gamma(\varE) \to \Gamma(\varE)$ such that
\begin{align}
\label{eq-definition-rules-nabla}
\nabla_X(f \psi)&= (X\cdotaction f) \psi + f \nabla_X \psi,
&
\nabla_{X+Y} \psi &= \nabla_X \psi + \nabla_Y \psi,
&
\nabla_{f X} \psi &= f \nabla_X \psi.
\end{align}
These relations are sufficient to define a covariant derivative on the space of smooth sections of any vector bundle $\varE$. The quantity $[\nabla_X, \nabla_Y] - \nabla_{[X,Y]}$ is a $C^\infty(\varM)$-linear map $\Gamma(\varE) \to \Gamma(\varE)$ which is the multiplication by $\gF(X,Y)$ (modulo a representation).

Consider a gauge transformation $\Xi$. Then the theory tells us that it induces an invertible map $\sigma : \Gamma(\varE) \to \Gamma(\varE)$ such that $\sigma (f \psi) = f \sigma(\psi)$ for any $f\in C^\infty(\varM)$. The covariant derivative $\nabla^\sigma$ associated to $\Xi^\pbast \omega$ is then given by $\nabla^\sigma_X \psi = \sigma^{-1} \circ \nabla_X \circ \sigma (\psi)$.

Using a trivialization of $\varE$ (induced by a trivialization $(\varU, \phi)$ of $\varP$), the section $\psi$ is a map $\varphi : \varU \to \evE$, and the covariant derivative takes the form
\begin{equation*}
D_X \varphi = X\cdotaction \varphi + \eta(A(X)) \varphi.
\end{equation*}
where $\eta$ is the representation of $\kg$ on $\evE$ induced by $\ell$. The action of a gauge group element $\gamma : \varU \to G$ on $\varphi$ is given by $\varphi^\gamma = \ell(\gamma)^{-1} \varphi$. In order to simplify notations, let us omit the representations $\ell$ and $\eta$ in the following.

\smallskip
Using a local coordinate system $(x^\mu)$ on $\varU$, this defines the differential operator $D_\mu = \partial_\mu + A_\mu$ where $D_\mu = D_{\partial_\mu}$ and $A_\mu = A(\partial_\mu) \in \kg$. This is the ordinary covariant derivative used in physics, which gives rise, in Lagrangians, to the minimal coupling ``$A_\mu \varphi$''. The field $\varphi$ supports the representation $\varphi \mapsto \gamma^{-1} \varphi$, and from a physical point of view, this is its characterization as a gauge field. Then the operator $\partial_\mu$ does not respect this representation, because $\gamma$, being ``local'' (it depends on $x \in \varU$), one has $\partial_\mu \varphi^\gamma = (\partial_\mu \gamma^{-1}) \varphi + \gamma^{-1} \partial_\mu \varphi = \gamma^{-1} (\partial_\mu  + (\gamma \partial_\mu \gamma^{-1}))  \varphi$, where in the last expression we make apparent a well defined object $\gamma \partial_\mu \gamma^{-1}$ with values in $\kg$, so that this last expression has a general meaning. On the contrary, the differential operator $D_\mu$ respects the representation, since $D_\mu \varphi^\gamma = (D_\mu \varphi)^\gamma = \gamma^{-1} D_\mu \varphi$. This is the heart of the usual formulation of gauge field theories: promote a global symmetry ($\gamma$ constant) to a local symmetry ($\gamma$ a function) by replacing $\partial_\mu$ everywhere in the Lagrangian by a differential operator $D_\mu$ which is compatible with the action of the gauge group. This requires to add new fields $A_\mu$ in the game with gauge transformations \eqref{eq-action-gauge-group-A-F}. But one needs also to introduce a gauge invariant term which describes the dynamics of the fields $A_\mu$. The simplest solution is the so-called Yang-Mills action
\begin{equation}
\label{eq-action-YM}
\Act_\text{Gauge, YM}[A] = \tfrac{1}{2} \int \tr(F \wedge  \hodgeast F) = \tfrac{1}{2} \int \tr(F_{\mu\nu} F^{\mu\nu}) \, \dvol
\end{equation}
where $\dvol$ is a metric volume element on $\varM$,
\begin{equation}
\label{eq-Fmunu}
F_{\mu \nu} = \partial_\mu A_\nu - \partial_\nu A_\mu + [A_\mu, A_\nu]
\end{equation}
is the local expression of the curvature $\Omega$, and $\tr$ is a Killing metric on $\kg$.

Notice that \eqref{eq-action-YM} is not the only admissible action functional for the fields $A$. The Chern-Simons action can be defined for space-times of dimension $3$ as
\begin{equation*}
\Act_\text{Gauge, CS}[A] = \int \tr\left( A \wedge \dd A + \tfrac{2}{3} A \wedge A \wedge A \right)
\end{equation*}
when the structure group $G$ is non abelian. This action is not gauge invariant, and only $e^{i \kappa \Act_\text{Gauge, CS}[A]}$ can be made gauge invariant by a suitable choice of $\kappa$.

\smallskip
Let us consider a more formal point of view about connections, which will be the key for generalizations of this notion in \ref{sec Basic structures in NCG} and \ref{sec Gauge field theories}. The Serre-Swan theorem \cite{Serr55a, Swan62a} tells us that a vector bundle $\varE$ on a smooth manifold $\varM$ is completely characterized by its space of smooth sections $\modM = \Gamma(\varE)$, which is a projective finitely generated module over the (commutative) algebra $C^\infty(\varM)$. Then, the assignment $X \mapsto \nabla_X$ defines a map
\begin{equation}
\label{eq-nabla-comm-module}
\nabla : \modM \rightarrow \Omega^1(\varM) \otimes_{\algA} \modM,
\quad\text{such that}\quad
\nabla (f \psi) = \dd f \otimes \psi + f \nabla \psi.
\end{equation}
This map can be naturally extended into a map
\begin{equation*}
\nabla : \Omega^\grast(\varM) \otimes_{\algA} \modM \rightarrow \Omega^{\grast+1}(\varM) \otimes_{\algA} \modM,
\end{equation*}
by the derivation rule
\begin{equation*}
\nabla (\eta \otimes \psi ) = \dd \eta \otimes \psi  + (-1)^{r} \eta \wedge \nabla \psi,  \quad \text{for any $\eta \in \Omega^r(\varM)$.}
\end{equation*}
The curvature of $\nabla$ is then defined as $\nabla^2 : \modM \rightarrow \Omega^{2}(\varM) \otimes_{\algA} \modM$, and it can be shown that $\nabla^2 (f \psi) = f \nabla^2 \psi$, and also that $\nabla^2 \psi$ is the multiplication by $\gF$ (modulo the representation $\eta$ of $\kg$ on $\evE$ mentioned before). The gauge transformation $\sigma : \modM \to \modM$ can be extended as an invertible map $\sigma : \Omega^\grast(\varM) \otimes_{\algA} \modM \to \Omega^\grast(\varM) \otimes_{\algA} \modM$, and one has 
\begin{equation}
\label{eq-action-sigma-nabla}
\nabla^\sigma = \sigma^{-1} \circ \nabla \circ \sigma.
\end{equation}

From the three levels of description of connections given above, it is clear that the covariant derivative $\nabla$ is related to the last one (``Globally on $\varM$'') because it acts on section of $\varE$ (which are globally defined on $\varM$). The differential operator $D$ corresponds to the second one, and there is a third description (not given here) which makes use of (equivariant) maps defined on $\varP$.

\subsection{Linear connections and Einstein theory of gravitation}
\label{sec linear connections}

As a vector bundle over $\varM$, the tangent bundle $T\varM$ is an essential structure for studying $\varM$ and its differential geometry. The global topology of this bundle is not arbitrary as could be the topology of the vector bundles considered above, and, as an associated vector bundle, it is related to the topology of the principal fiber bundle $L\varM$ of frames on $\varM$. 

Then, the theory of connections defined in this situation is quite different from the one defined above on arbitrary principal fiber bundle. A linear connection is a connection defined on $L\varM$. It is usual to look at this connection as a covariant derivative $\nabla_X : \Gamma(T\varM) \to \Gamma(T\varM)$ which satisfies \eqref{eq-definition-rules-nabla}. In addition to the curvature defined as $[\nabla_X, \nabla_Y] - \nabla_{[X,Y]}$, it is possible to introduce a $2$-form in $\Omega^2(\varM, T\varM)$ which is specific to linear connections, the torsion $T(X,Y) = \nabla_X Y - \nabla_Y X - \nabla_{[X,Y]} \in \Gamma(T\varM)$. Another singular and important object is the  soldering form $\theta \in \Omega^1(\varM, T\varM)$ defined by $\theta(X) = X$ for any $X\in \Gamma(T\varM)$. The torsion is related to $\theta$ by $T = \nabla \theta$.

Following what have been explained in \ref{sec Ehresmann connections}, let us consider $L\varM$ as a principal fiber bundle with structure group $GL(n,\gR)$, and let us introduce a connection $\omega$ on it. Then this connection induces covariant derivatives on any associated vector bundles, for instance $\nabla$ on $T\varM$, but also on the bundle of forms, and generally on any bundle of tensors on $\varM$. Doing that, the gauge group is defined as $\caG(L\varM)$, so that, locally, a gauge transformation is a map $\gamma : \varU \to GL_n(\gR)$.

It is also natural to look locally at the covariant derivative $\nabla$ using a coordinate system $(x^\mu)$ on an open subset $\varU \subset \varM$. Then the local derivations $\partial_\mu$ induce natural basis on each tangent space over $\varU$, and, using \eqref{eq-definition-rules-nabla}, $\nabla$ is completely determined by the quantities $\Gamma_{\mu \nu}^\rho$, the Christoffel symbols, defined by
\begin{equation*}
\nabla_{\partial_\mu} \partial_\nu = \Gamma_{\mu \nu}^\rho \partial_\rho.
\end{equation*}
The $\Gamma_{\mu \nu}^\rho$'s define the local trivialization of $\omega$ if one uses the $\partial_\mu$ as a local trivialization of $L\varM$. Straightforward computations then give the curvature and the torsion as
\begin{align*}
{R^\rho}_{\sigma\, \mu\nu} &= \partial_\mu \Gamma_{\nu \sigma}^\rho - \partial_\nu \Gamma_{\mu \sigma}^\rho + \Gamma_{\mu \eta}^\rho \Gamma_{\nu \sigma}^\eta - \Gamma_{\nu \eta}^\rho \Gamma_{\mu \sigma}^\eta,
&
T_{\mu \nu}^\rho &= \Gamma_{\mu \nu}^\rho - \Gamma_{\nu \mu}^\rho.
\end{align*}
The expression of the curvature is \eqref{eq-Fmunu} in this specific situation. The Ricci tensor is then defined as a contraction of the curvature: $R_{\sigma \nu} = {R^\rho}_{\sigma\, \rho\nu}$.

The Christoffel symbols determine completely the linear connection $\omega$, so defining $\nabla$ is sufficient to introduce a linear connection. In the following we will identify a linear connection with its covariant derivative on $\Gamma(T\varM)$.

\medskip
Let us now introduce a metric $g$ on $\varM$.  A linear connection $\nabla$ is said to be metric if it satisfies $X\cdotaction g(Y,Z) = g(\nabla_X Y, Z) + g(Y, \nabla_X Z)$ for any $X, Y, Z \in \Gamma(T\varM)$. It is well-known that there is a unique torsionless metric linear connection $\nabla^\LC$. It is the Levi-Civita connection, whose Christoffel symbols are
\begin{equation}
\label{eq-Christoffel-LC}
\Gamma_{\mu \nu}^\rho = \tfrac{1}{2} g^{\rho\sigma}\left( \partial_\nu g_{\sigma \mu} + \partial_\mu g_{\sigma \nu} - \partial_\sigma g_{\mu\nu} \right).
\end{equation}
The metric $g$ can be used to contract indices of the Ricci tensor $R_{\sigma \nu}$ to produce the scalar curvature $R = g^{\sigma\nu}R_{\sigma \nu}$.

The Einstein's theory of gravitation has been formulated historically in terms of these structures, using the action
\begin{equation}
\label{eq-EH}
\Act[g] = \frac{-1}{16 \pi G} \int R \, \sqrt{\abs{g}}\dd^4 x. 
\end{equation}
on a $4$-dimensional space-time manifold $\varM$.

Written in this form, \textsl{i.e.} starting from the metric $g$ as the primary field, this theory is not a gauge field theory. Since the Lagrangian $\caL = R$ is a scalar in terms of natural structures on $\varM$ (see \cite{KolaMichSlov93a} for this notion), the theory is only invariant under the action of the group of diffeomorphisms of $\varM$: there is no gauge group in the theory, in the sense of  \ref{sec Ehresmann connections}. This is also confirmed by the fact that the primary object in the theory is the metric field $g$, and not a connection $1$-form $\omega$, which derives from $g$ by \eqref{eq-Christoffel-LC}. The covariant derivative is a byproduct, whose purpose is for instance to write equations of motion of point-like objects, in the form of geodesic equations.

Nevertheless, if one wants to stress the importance of the linear connection $\nabla^\LC$, one is tempted to look at this theory as a gauge theory for the gauge group $\caG(L\varM)$, which is related to the diffeomorphisms group of $\varM$ by the short exact sequence:
\begin{equation}
\label{eq-ses-aut-LM}
\xymatrix@1{{\algun} \ar[r] & {\caG(L\varM)} \ar[r] & {\Aut(L\varM)}  \ar[r] & {\Diff(\varM)} \ar[r] & {\algun}},
\end{equation}
where $\Aut(L\varM)$ is the group of all automorphisms of the principal fiber bundle $L\varM$. But this group $\Aut(L\varM)$ is superfluous, since all the symmetries of $\Act[g]$ are already in the group  $\Diff(\varM)$.

\subsection{Cartan geometry and gravitation theories}
\label{sec Cartan geometry}

Provided one departs from the theory of connections described in \ref{sec Ehresmann connections}, it is possible to write Einstein's theory of gravitation as a gauge field theory.

The original idea of Klein, formulated in his \textsl{Erlangen program} of 1872, is to characterize a geometry as the study of the invariants of an homogeneous and isotropic space. In modern language, a Klein geometry is a couple of Lie groups $(G, H)$ such that $H$ is a closed subgroup of $G$, and $G/H$ is the homogeneous space.

The purpose of Cartan geometry is to consider a global manifold which can be locally modeled on a Klein geometry $(G,H)$. In the following, we use the bundle definition of a Cartan geometry, described as follows \cite{Shar97a}. 

Let $(G, H)$ be as before, and denote by $(\kg, \kh)$ the associated Lie algebras. A Cartan geometry consists in the following data:
\begin{enumerate}
\item a principal fiber bundle $\varP$ with structure group $H$ on a smooth manifold $\varM$;

\item a $\kg$-valued $1$-form $\varpi$ on $\varP$ such that:
\begin{enumerate}
\item ${\tR}_h^\pbast \varpi = \Ad_{h^{-1}}\varpi$ for any $h \in H$,

\item $\varpi(\xi^\varP) = \xi $ for any $\xi \in \kh$,

\item at each point $p \in \varP$, the linear map $\varpi_p : T_p \varP \to \kg$ is an isomorphism.
\label{item-iso}
\end{enumerate}

\end{enumerate}
Notice that the dimension of $G$ is exactly the dimension of $\varP$, \textsl{i.e.} the dimension of $\varM$ plus the dimension of $H$. Condition \ref{item-iso} is a strong requirement: the principal fiber bundle $\varP$ is ``soldered'' to the base manifold, which constrains its global topology.

The curvature $\bOmega \in \Omega^2(\varP) \otimes \kg$ is defined as $\bOmega = \dd \varpi + \tfrac{1}{2} [\varpi, \varpi]$, it vanishes on vertical vector fields. Denote by $\rho : \kg \to \kg/\kh$ the quotient map, then the torsion is $\rho(\bOmega)$. For any $\xi \in \kg$ and $p \in \varP$, $\varpi_p^{-1}(\xi) \in T_p\varP$. But if $\xi \in \kh$, then $\varpi_p^{-1}(\xi) \in V_p\varP$ (vertical vectors in $T_p\varP$), so that $\bOmega_p(\varpi_p^{-1}(\xi), X_p) = 0$ for any $X_p \in T_p \varP$.

A simple example consists to consider $\varP = G$, $\varM = G/H$ and $\varpi = \theta_G$, the Maurer-Cartan $1$-form on $G$. Then the curvature is zero. This is the Klein model on which general Cartan geometries are based.

Let $\Xi \in \caG(\varP)$ be a vertical automorphism of $\varP$. Then it acts by pull-back on $\varpi$ and $(\varP, \Xi^\pbast \varpi)$ defines another Cartan geometry. The gauge group of a Cartan geometry is then the (ordinary) gauge group of $\varP$.

\smallskip
A reductive Cartan geometry corresponds to a situation when one has a $H$-module decomposition $\kg = \kh \oplus \kp$, where $\kp$ is a $\Ad_H$-module (reductive decomposition of $\kg$). Then the Cartan connection $\varpi$ splits as $\varpi = \omega \oplus \beta$, where $\omega$ takes its values in $\kh$ and $\beta$ in $\kp$. From the hypothesis on $\varpi$, one can check that $\omega$ is an Ehresmann connection on $\varP$. The $1$-form $\beta$ is called a soldering form on $\varP$: for any $p \in \varP$, it realizes an isomorphism $\beta_p : H_p \varP \to \kp$, where $H_p\varP = \ker \omega_p$ is the horizontal subspace of $T_p\varP$ associated to $\omega$. In particular, it vanishes on $V_p\varP$. The curvature $2$-form splits as well into  $\bOmega = \Omega + \tau(\bOmega)$, where $\Omega$ is the curvature of $\omega$ in the sense of \ref{sec Ehresmann connections}, and $\tau(\bOmega)$ is the torsion. 

Let $(\varU, \phi)$ be a local trivialization of $\varP$, and let $s : \varU \to \varP_{|\varU}$ be its associated local section. Denote by $\Gamma = s^\pbast \omega$ and $\Lambda = s^\pbast \beta$ the local trivializations of $\omega$ and $\beta$. Then $\Lambda_x : T_x \varU \to \kp$ is an isomorphism. Let $\eta$ be a $\Ad_H$-invariant bilinear form on $\kp$. Then, for any $X_1, X_2 \in T_x \varU$, $g_x(X_1, X_2) = \eta(\Lambda_x(X_1),\Lambda_x(X_2))$ defines a metric on $\varM$. Using local coordinates $(x^\mu)$ on $\varU$, and a basis $\{e_a\}$ of $\kp$, one has $\Lambda = {\Lambda^a}_\mu \dd x^\mu \otimes e_a$, and $g_{\mu\nu} = \eta_{ab} {\Lambda^a}_\mu {\Lambda^b}_\nu$ with obvious notations. This relation is well known in the tetrad formulation of General Relativity.

\smallskip
Consider now a reductive Cartan geometry, on an orientable manifold $\varM$, based on the groups $G = SO(1, m-1) \ltimes \gR^{m}$ and $H = SO(1, m-1)$, so that $\kp =  \gR^{m}$. Let $\varpi = \omega \oplus \beta$ be a Cartan connection on $\varP$, and denote as before $\Gamma$ and $\Lambda$ the local trivializations of $\omega$ (the spin connection) and $\beta$. Let us introduce a basis $\{e_a\}_{1 \leq a \leq m}$ of $\gR^{m}$, so that any element $\xi \in \kh$ is a matrix $({\xi^a}_b)_{1 \leq a, b \leq m}$. The local $1$-form $\Gamma$ can be written as $\Gamma = \left( {\Gamma^a}_{b\mu} \dd x^\mu \right)_{1 \leq a, b \leq m}$ and the local $1$-form $\Lambda$ is vector-valued $\left( {\Lambda^a}_\mu \dd x^\mu \right)_{1 \leq a \leq m}$. Denote by $R = \left( {R^a}_{b\mu\nu} \dd x^\mu \wedge \dd x^\nu \right)_{1 \leq a, b \leq m}$ the local expression of the curvature of $\omega$. Finally, define $g$ as the metric on $\varM$ induced by $\beta$ and the $\Ad_H$-invariant Minkowski metric $\eta$ on $\gR^m$, and denote by $\hodgeast$ its Hodge star operator.

The action functional of General Relativity can be written as $\Act_\text{Gauge} + \Act_\text{Matter}$ where
\begin{equation}
\label{eq-gauge-grav}
\Act_\text{Gauge}[\omega, \beta] = \frac{-1}{32 \pi G}\int R^a{}_b \wedge \hodgeast(\Lambda^b \wedge \Lambda_a) 
\end{equation}
is the (tetradic) Palatini action functional and $\Act_\text{Matter}$ is the action functional of the matter, which depends also on $\beta$ and $\omega$ by minimal coupling. The equations of motion are obtained by varying $\beta$ and $\omega$ independently: the first one gives the usual Einstein's equations, and varying the spin connection relates the torsion $\tau(\bOmega)$ to the spin of matter (see \cite{GockSchu89a} for details). When the spin of matter is zero, the torsion vanishes, and one gets the usual Einstein's theory of gravitation. The model described above is the Einstein-Cartan version of General Relativity, which takes into account spin of matter.

In a reductive Cartan geometry, the principal fiber bundle $\varP$ is necessarily a reduction of the  $GL_n^+(\gR)$-principal fiber bundle $L\varM$ to the subgroup $H$ \cite[Lemma~A.2.1]{Shar97a}. In the present case, the soldering form realizes an isomorphism between $\varP$ and the $SO(1, m-1)$-principal fiber bundle of orthonormal frames on $\varM$ for the metric $g$ (induced by $\beta$). The usual point of view is to consider that the theory defined on $\varP$ is induced, through a metric $g$, by a symmetry reduction $GL_n^+(\gR) \to SO(1, m-1)$, in which $\omega$ is obtained from a $g$-compatible linear connection $\nabla$ on $L\varM$ \cite{Trau79a}.

But, following the general scheme presented in~\cite{FourFranLazzMass14a}, one can consider the linear connection $\nabla$ on $L\varM$ as the result of the decoupling of the gauge symmetry on $\varP$ to ``nothing''. Indeed, from the local trivializations $\Gamma$ and $\Lambda$, one can construct the gauge invariant composite fields $\Lambda^{-1} \Gamma\Lambda + \Lambda^{-1}\dd \Lambda$, which turn out to behave geometrically as the Christoffel symbols of a linear connection $\nabla$ on $L\varM$. This amounts to decoupling completely the $SO(1, m-1)$ internal gauge degrees of freedom, and the result is a purely geometric theory. Accordingly, the gauge invariant Einstein-Cartan  action~\eqref{eq-gauge-grav} reduces to the geometrically well-defined Einstein-Hilbert action~\eqref{eq-EH} constructed on tensorial quantities only. This procedure can be interpreted with the help of the short exact sequence of groups
\begin{equation}
\label{eq-ses-Cartan}
\xymatrix@1{{\algun} \ar[r] & {SO(1, m-1)} \ar[r] & {SO(1, m-1)\ltimes \gR^m}  \ar[r] & {\gR^m} \ar[r] & {\algzero}},
\end{equation}
where the decoupling of the $SO(1, m-1)$ part of the symmetry reduces the total symmetry based on $SO(1, m-1)\ltimes \gR^m$ to the diffeomorphisms only (encoded in the $\gR^m$ part).

Thus, the original geometric formulation of Einstein's theory of gravitation can be \emph{lifted} to a gauge field theory in the framework of reductive Cartan geometries. But this construction does not make apparent new fundamental symmetries: following the procedure introduced in~\cite{FourFranLazzMass14a}, a full decoupling of the gauge group can be realized as a mere change of variables in the space of fields, and it gives rise to the original formulation of the theory.

\section{\NC geometry}
\label{sec-NC}

\NCG is not a physical theory, contrary to string theories or quantum loop gravity. It a mathematical research activity which has emerged in the 80's \cite{Conn85a, Conn00a, ConnMarc08a, Dubo01a, Dubo91a, Land97a, GracVariFigu01a} at the intersection of differential geometry, normed algebras and representation theory. In particular, as a generalization of ordinary differential geometry, \ncg has shed new lights on gauge field theories. A notion of connections can be defined in terms of modules and differential calculi, which is the natural language of \ncg.

The main difficulty to get a clear view of these achievements comes from the fact that many approaches have been proposed to study the differential structure of \nc spaces. Two of them will be of interest here. The theory of spectral triples, developed by Connes, emphasizes the metric structure \cite{Conn94a, ConnMarc08a, GracVariFigu01a}, which is encoded into a Dirac operator. On the other hand, many \nc spaces can be studied through a more canonical differential structure \cite{Coqu92a, Coqu92b, CoquEspoVail91a, Dubo91a, Dubo97a, Dubo01a, Mass08b, Mass12a, Mass08c, CagnMassWall11a, Mass99a, Mass96a}, based on the space of derivations of associative algebras. 

In spite of that, many of the \nc gauge field theories that have been developed and studied so far use essentially the same ideas and the same building blocks. Independently of their exact constitutive elements, many of these gauge theories share some common or similar features, among them the origin of the gauge group and the possibility to naturally produce Yang-Mills-Higgs Lagrangians.

The following review on gauge field theories in \ncg can be completed by~\cite{Mass12a}.

\subsection{Basic structures in \ncg}
\label{sec Basic structures in NCG}

\NCG relies on fundamental theorems which identify the good algebra of functions on specific spaces that encode all the structure of the space. For instance, the Gelfand-Naïmark theorem tells us that a unital commutative $C^\ast$-algebra is always the commutative algebra of continuous functions on a compact topological space, equipped with the sup norm. Studying a commutative $C^\ast$-algebra is studying the underlying topological space. In the same way, one can study a measurable space using the commutative von~Neumann algebra of bounded measurable functions. For differentiable manifolds, no such theorem has been established. Nevertheless, the reconstruction theorem by Connes \cite{Conn08a} is an attempt to characterize spin manifolds using commutative spectral triples.

These fundamental theorems do not only identify the good (category of) algebras, they also produce a collection of tools to study these spaces using only these algebras. And these tools are defined on, or can be generalized to, \nc algebras in the same category. Among the fundamental tools, two of them must be mentioned: $K$-theory \cite{RordLarsLaus00a, HigsRoe04a, Wegg93a, Blac98a}, and it dual $K$-homology which is at the heart of the mathematical motivation for spectral triples, and cyclic homology \cite{CuntKhal97a, Loda98a}. These tools permits to revoke the assumption about the commutativity of the algebra describing the space under study, and to consider ``\nc'' versions of these spaces as \nc algebras in the same category.

The theory of vector bundles plays an essential role on gauge field theories, as mentioned in Section~\ref{sec-Ordinary differential geometry}. The good \nc notion of vector bundle is played by projective finitely generated modules over the algebra. This characterization relies on theorems by Serre and Swan \cite{Serr55a, Swan62a} which identify in this algebraic way sections of vector bundles.

Connections use also some notion of differentiability, for instance the de~Rham differential or the covariant derivative. In \ncg, it is accustomed to replace the de~Rham space of forms by a differential calculus associated to the associative algebra we want to study. In the following, we will assume that the reader is familiar with certain basic algebraic notions, such as associative algebras, modules, graduations, involutions (see \cite{Jaco85a} for instance).

A differential calculus on an associative algebra $\algA$ is a graded differential algebra $(\Omega^\grast, \dd)$ such that $\Omega^0 = \algA$. The space $\Omega^p$ is called the space of \nc $p$-forms (or $p$-forms in short), and it is automatically a $\algA$-bimodule. By definition, $\dd : \Omega^{\grast} \to \Omega^{\grast+1}$ is a linear map which satisfies $\dd (\omega_p \eta_q) =  (\dd \omega_p) \eta_q + (-1)^p \omega_p (\dd \eta_q)$ for any $\omega_p \in \Omega^p$ and $\eta_q \in \Omega^q$. We will suppose that $\algA$ has a unit $\bbbone$, and then this property implies $\dd \bbbone = 0$. When $\algA$ is equipped with an involution $a \mapsto a^\invast$, we can suppose that the graded algebra $\Omega^\grast$ has also an involution, denoted by $\omega_p \mapsto \omega_p^\invast$, which satisfies $(\omega_p \eta_q)^\invast = (-1)^{pq} \eta_q^\invast \omega_p^\invast$ for any $\omega_p \in \Omega^p$ and $\eta_q \in \Omega^q$, and we suppose that the differential operator $\dd$ is real for this involution: $(\dd \omega_p)^\invast = \dd (\omega_p^\invast)$.

There are many ways to define differential calculi, depending on the algebra under investigation. Two of them will be of great interest in the following. The first one is the de~Rham differential calculus $(\Omega^\grast(M), \dd)$ on the algebra $\algA = C^\infty(\varM)$, where $\varM$ is a smooth manifold. This one need not be described further. It is the ``commutative model'' of \nc geometry. 

The second one can be attached to any unital associative algebra: it is the universal differential calculus, denoted by $(\Omega^\grast_U(\algA), \dd_U)$ (see for instance \cite{Dubo97a} for a concrete construction). It is defined as the free unital graded differential algebra generated by $\algA$ in degree $0$. The unit in $\Omega^\grast_U(\algA)$ is also a unit for $\Omega^0_U(\algA) = \algA$, so that it coincides with the unit $\bbbone$ of $\algA$. This differential calculus has an universal property (so its name) formulated as follows: for any unital differential calculus $(\Omega^\grast, \dd)$ on $\algA$, there exists a unique morphism of unital differential calculi $\phi : \Omega^\grast_U(\algA) \rightarrow \Omega^\grast$ (of degree $0$) such that $\phi(a) = a$ for any $a\in \algA = \Omega^0_U(\algA) = \Omega^0$. This universal property permits to characterize all the differential calculi on $\algA$ generated by $\algA$ in degree $0$ as quotients of the universal one. Even if $\algA$ is commutative, $\Omega^\grast_U(\algA)$ need not be graded commutative. 

An explicit construction of $(\Omega^\grast_U(\algA), \dd_U)$ describes $\Omega^n_U(\algA)$ as finite sum of elements $a \dd_U b_1 \cdots \dd_U b_n$ for $a, b_1, \dots, b_n \in \algA$, where the notation $\dd_U b$ can be considered as formal, except that it takes into account the important relation $\dd_U \bbbone = 0$. Then if $\algA$ is involutive, the involution on $\Omega^\grast_U(\algA)$ is defined as
\begin{equation*}
(a \dd_U b_1 \cdots \dd_U b_n)^\invast = (-1)^{\tfrac{n(n-1)}{2}} (\dd_U b_n^\invast) \cdots (\dd_U b_1^\invast ) a^\invast.
\end{equation*}
This differential calculus is strongly related to Hochschild and cyclic homology \cite{CuntKhal97a, Loda98a, Mass08d}.

\medskip
\NC connections are defined using the characterization \eqref{eq-nabla-comm-module} of ordinary connections, and the fact that, due to the Serre-Swan theorem, the good notion of ``\nc vector bundle'' is the notion of (projective finitely generated) module.

Let $\modM$ be a right $\algA$-module, and let $(\Omega^\grast, \dd)$ be a differential calculus on $\algA$. Then a \nc connection on $\modM$ is a linear map 
\begin{equation}
\label{eq-hatnabla-comm-module}
\hnabla : \modM \rightarrow \modM \otimes_\algA \Omega^1,
\quad\text{such that}\quad
\hnabla(ma) = (\hnabla m) a + m \otimes \dd a,
\end{equation}
for any $m \in \modM$ and $a \in \algA$. This map can be extended as $\hnabla : \modM \otimes_\algA \Omega^p \rightarrow \modM \otimes_\algA \Omega^{p+1}$, for any $p \geq 0$, using the derivation rule 
\begin{equation*}
\hnabla (m \otimes \omega_p) = (\hnabla m) \otimes \omega_p + m \otimes \dd \omega_p
\quad\text{for any $\omega_p \in \Omega^p$.}
\end{equation*}
The curvature of $\hnabla$ is then defined as $\hR = \hnabla^2 = \hnabla \circ \hnabla : \modM \rightarrow \modM \otimes_\algA \Omega^2$, and it satisfies $\hR (ma) = (\hR m) a$ for any $m \in \modM$ and $a \in \algA$. The space $\caA(\modM)$ of \nc connections on $\modM$ is an affine space modeled on the vector space $\Hom^\algA(\modM,\modM \otimes_\algA \Omega^1)$ of right $\algA$-modules morphisms.

Suppose now that $\algA$ has an involution. A Hermitian structure on $\modM$ is a $\gR$-bilinear map $\langle -, - \rangle : \modM \otimes \modM \rightarrow \algA$ such that $\langle ma, nb \rangle = a^\invast \langle m, n \rangle b$ and $\langle m, n \rangle^\invast = \langle n, m \rangle$ for any $a,b \in \algA$ and $m,n \in \modM$. There is a natural extension $\langle -, - \rangle$ to $(\modM \otimes_\algA \Omega^p) \otimes (\modM \otimes_\algA \Omega^q) \rightarrow \Omega^{p+q}$ defined by $\langle m \otimes \omega_p, n \otimes \eta_q \rangle = \omega_p^\invast \langle m, n \rangle \eta_q$. A \nc  connection $\hnabla$ is said to be compatible with $\langle -, - \rangle$, or Hermitian, if, for any $m,n\in \modM$,
\begin{equation*}
\langle \hnabla m, n \rangle + \langle m, \hnabla n \rangle = \dd \langle m, n \rangle.
\end{equation*}

In this context, the gauge group $\caG$ is then defined as the group of automorphisms of $\modM$ as a right $\algA$-module: $\Phi \in \caG$ satisfies $\Phi(ma) = \Phi(m)a$ for any $m \in \modM$ and $a \in \algA$. It depends on the choice of the right module $\modM$. We can extend a gauge transformation $\Phi$ to a right $\Omega^\grast$-module automorphism on $\modM \otimes_\algA \Omega^\grast$ by $\Phi(m \otimes \omega) = \Phi(m) \otimes \omega$. Generalizing \eqref{eq-action-sigma-nabla}, we can show that the map
\begin{equation*}
\hnabla^\Phi = \Phi^{-1} \circ \hnabla \circ \Phi
\end{equation*}
is a \nc  connection on $\modM$. This defines the action of gauge transformation on $\caA(\modM)$.

A gauge transformation $\Phi$ is said to be compatible with the Hermitian structure $\langle -, - \rangle$ if $\langle \Phi(m), \Phi(n) \rangle = \langle m, n \rangle$ for any $m,n \in \modM$. Denote by $U(\caG)$ the subgroup of $\caG$ of gauge transformations which preserve $\langle -, - \rangle$. This subgroup defines an action on the subspace of \nc connections compatible with $\langle -, - \rangle$.

A natural question is to ask if the space $\caA(\modM)$ is not empty. There is a natural condition on $\modM$ (suggested by the Serre-Swan theorem) which solves this problem. If $\modM$ is a projective finitely generated right module, then $\caA(\modM)$ is not empty. Indeed, the condition means that there is an integer $N >0$ and a projection $p \in M_N(\algA)$ such that $\modM \simeq p \algA^N$. Then $p$ extends to a map $(\Omega^\grast)^N \rightarrow (\Omega^\grast)^N$ which acts on the left by matrix multiplication and one has $\modM \otimes_\algA \Omega^\grast = p (\Omega^\grast)^N$. Let $\hnabla^0$ be a \nc connection on the right module $\algA^N$. Then it is easy to show that $m \mapsto p \circ \hnabla^0 m$ is a \nc connection on $\modM$, where $m \in \modM \subset \algA^N$. Notice then that $\hnabla^0 m = \dd m$ is a \nc connection on $\algA^N$, so that $\caA(\modM)$ is not empty. The associated \nc connection is given by $\hnabla m = p \circ \dd m$ on $\modM$, and its curvature is the left multiplication on $\modM \subset \algA^N$ by the matrix of $2$-forms $p \dd p \dd p$.

\smallskip
By construction, this definition of \nc connections is a direct generalization of covariant derivatives on associated vector bundles. There is a way to introduce algebraic structures (\nc forms) to replace this \nc covariant derivative. In order to simplify the presentation, we will consider the particular case $\modM = \algA$. See \cite{Mass12a} for the more general situation.

With $\modM = \algA$, one has $\modM \otimes_\algA \Omega^\grast = \Omega^\grast$, and since $\algA$ is unital, one has $\hnabla (a) =\hnabla (\bbbone a) = \hnabla(\bbbone) a + \bbbone \otimes \dd a = \hnabla(\bbbone) a + \dd a$. This implies that $\hnabla(\bbbone) = \omega \in \Omega^1$ characterizes completely $\hnabla$. We call $\omega$ the connection $1$-form of $\hnabla$, and the curvature of $\hnabla$ is the left multiplication by the $2$-form $\Omega = \dd \omega + \omega \omega \in \Omega^2$. An element $\Phi$ of the gauge group is completely determined by $\Phi(\bbbone) =g \in \algA$ (invertible element). It acts on $\modM$ by left multiplication: $\Phi(a) = g a$. A simple computation shows that the connection $1$-form associated to $\hnabla^\Phi$ is $\omega^g = g^{-1} \omega g + g^{-1} \dd g$ and its curvature $2$-form is $g^{-1}(\dd \omega + \omega \omega) g = g^{-1} \Omega g$. These relations can be compared to \eqref{eq-action-gauge-group-omega-Omega} or \eqref{eq-action-gauge-group-A-F}. When $\algA$ is involutive, $\langle a, b \rangle = a^\invast b$ defines a natural Hermitean structure on $\modM$, and one has $U(\caG) = U(\algA)$, the group of unitary elements in $\algA$.

\subsection{Spectral triples}
\label{sec Spectral triples}

In order to simplify the presentation, we will restrict ourselves to compact spectral triples, \textsl{i.e.} the algebras will be unital.

A spectral triple $(\algA, \ehH, \caD)$ is composed of a unital $C^\ast$-algebra $\algA$, a faithful involutive representation $\pi : \algA \rightarrow \caB(\ehH)$ on a Hilbert space $\ehH$, and an unbounded self-adjoint operator $\caD$ on $\ehH$, called a Dirac operator, such that:
\begin{itemize}
\item the set $\caA = \{ a \in \algA \ / \ [\caD, \pi(a)] \text{ is bounded} \}$ is norm dense in $\algA$;
\item $(1+\caD^2)^{-1}$ has compact resolvent.
\end{itemize}

The main points of this definition is that the representation makes $\ehH$ into a left $\algA$-module, and the Dirac operator $\caD$ defines a differential structure (more on this later). The sub algebra $\caA$ identifies with the ``smooth functions'' on the \nc space and the differential of $a \in \caA$ is more or less $\dd a = [\caD, a]$. Be aware of the fact that this can only be an heuristic formula since the commutator with $\caD$ cannot be used to define a true differential.

The spectral property of $\caD$ is used to define the dimension $n$ of the spectral triple through the decreasing rate of the eigenvalues of $|\caD|^{-1}$. The Dirac operator gives also a geometric structure to the spectral triple, in the sense that it gives a way to measure ``lengths'' between states. See \cite{Conn94a} for further details.

A spectral triple is said to be even when its dimension $n$ is even and when there exists a supplementary operator $\gamma : \caH \rightarrow \caH$ such that $\gamma^\invast = \gamma$, $\caD \gamma + \gamma \caD =0$, $\gamma \pi(a) - \pi(a) \gamma =0$, and $\gamma^2 = 1$, for any $a \in \algA$. This operator is called chirality.

A spectral triple is said to be real when there exists an anti-unitary operator $J : \caH \rightarrow \caH$ such that $[J \pi(a) J^{-1}, \pi(b)] =0$, $J^2 = \epsilon$, $J \caD = \epsilon' \caD J$ and $J \gamma = \epsilon'' \gamma J$ for any $a,b \in \algA$. The coefficients $\epsilon, \epsilon'$, and $\epsilon''$ take their values according to the dimension $n$ of the spectral triple as given in Table~\ref{table-epsilon}.

\begin{table}[t]
{\centering
\begin{tabular}{*{9}{>{$}r<{$}}}
\toprule
n \mod 8   & 0 &  1 &  2 &  3 &  4 &  5 &  6 & 7 \\
\midrule
\epsilon   & 1 &  1 & -1 & -1 & -1 & -1 &  1 & 1 \\
\epsilon'  & 1 & -1 &  1 &  1 &  1 & -1 &  1 & 1 \\
\epsilon'' & 1 &    & -1 &    &  1 &    & -1 &   \\
\bottomrule
\end{tabular}
\par}
\caption{Coefficients $\epsilon, \epsilon'$, and $\epsilon''$ according to the dimension $n$ of the spectral triple.}
\label{table-epsilon}
\end{table} 

By definition, $J \pi(a)^\invast J^{-1}$ commutes with $\pi(\algA)$ in $\caB(\ehH)$ (bounded operators on $\ehH$), so the involutive representation $a \mapsto J \pi(a)^\invast J^{-1}$ of $\algA$ on $\ehH$ induces a structure of $\algA$-bimodule on $\ehH$. We denote it by $(a,b) \mapsto \pi(a) J \pi(b)^\invast J^{-1}\Psi \simeq \pi(a) \Psi \cpi(b)$ for any $\Psi \in \ehH$ (the presence of $J$ in this formula implies the use of $\pi(b)^\invast$ instead of $\pi(b)$). Then the operator $\caD$ is required to be a first order differential operator for this bimodule structure \cite{DuboMass96b}: 
\begin{equation*}
\big[ [\caD, \pi(a)], J \pi(b) J^{-1}  \big] = 0\quad\text{for any $a,b \in \algA$.}
\end{equation*}

We have presented here a restricted list of axioms for a spectral triple, but it is sufficient to understand the principles of the gauge theories constructed in this approach.

Let us give a first example, which is the commutative model. Let $\varM$ be a smooth compact Riemannian spin manifold of dimension $m$, and let $\algA = C(\varM)$ be the commutative algebra of continuous functions on $\varM$. With the sup norm, this is a (commutative) $C^\ast$-algebra. Let $\slashed{S}$ be a spin bundle given by the spin structure on $\varM$, and let $\ehH = L^2(\slashed{S})$ be the associated Hilbert space. The Dirac operator $\caD = \slashed{\partial} = i \gamma^\mu \partial_\mu$ is the (usual) Dirac operator on $\slashed{S}$ associated to the Levi-Civita connection (spin connection in this context),  where the $\gamma^\mu$'s are the Dirac gamma matrices satisfying $\{\gamma^\mu, \gamma^\nu\} = 2 g^{\mu \nu}$. The dimension of the spectral triple $(\algA, \ehH, \caD)$ is $m$, and the sub algebra $\caA$ is $C^\infty(\varM)$. When $m$ is even, the chirality is given by $\gamma_\varM = - \gamma^1 \gamma^2 \cdots \gamma^m$. The charge conjugation defines a real structure $J_\varM$ on this spectral triple.

\smallskip
We will say that two spectral triples $(\algA, \ehH, \caD)$ and $(\algA', \ehH', \caD')$ are unitary equivalent if there exists a unitary operator $U : \ehH \rightarrow \ehH'$ and an algebra isomorphism $\phi : \algA \rightarrow \algA'$ such that $\pi' \circ \phi = U \pi U^{-1}$, $\caD' = U \caD U^{-1}$, $J' = U J U^{-1}$, and $\gamma' = U \gamma U^{-1}$ (the last two relations are required only when the operators $J$, $J'$, $\gamma$ and $\gamma'$ exist). 

Then we define a symmetry of a spectral triple as a unitary equivalence between two spectral triples such that $\ehH' = \ehH$, $\algA' = \algA$, and $\pi' = \pi$. In that case, $U : \ehH \rightarrow \ehH$ and $\phi \in \Aut(\algA)$. A symmetry acts only on the operators $\caD$, $J$ and $\gamma$. Let us consider the symmetries for which the automorphisms $\phi$ are $\caA$-inner: there is a unitary $u \in U(\caA)$ (unitary elements in $\caA$) such that $\phi_u(a) = u a u^\invast$ for any $a \in \algA$. Such a unitary defines all the symmetry, with $U = \pi(u) J \pi(u) J^{-1} : \ehH \rightarrow \ehH$. Considering the bimodule structure on $\ehH$, $U$ is the conjugation with $\pi(u)$: $\pi(u) J \pi(u) J^{-1} \Psi \simeq \pi(u) \Psi \cpi(u)^\invast$. A direct computation shows that inner symmetries leave invariant $J$ and $\gamma$, while the operator $\caD$ is modified as
\begin{equation}
\label{eq-Du}
\caD^u = \caD + \pi(u) [\caD, \pi(u)^\invast] + \epsilon' J\left( \pi(u) [\caD, \pi(u)^\invast] \right) J^{-1}.
\end{equation}

We define a gauge transformation as a unitary $u \in U(\caA)$ which acts on the spectral triple as defined above. This looks different from the definition proposed in \ref{sec Basic structures in NCG}, but we will show how the two points of view can be reconciled.

As in ordinary differential geometry, the ordinary derivative, here played by $\caD = i\gamma^\mu \partial_\mu$, is not invariant by gauge transformations, and we need an extra field to compensate for the inhomogeneous terms in \eqref{eq-Du}. Differential forms $\sum_i a_i \dd_U b^1_i \cdots \dd_U b^n_i$ in the universal differential calculus $(\Omega^\grast_U(\caA), \dd_U)$ defined in \ref{sec Basic structures in NCG} can be represented on $\ehH$ as
\begin{equation*}
\pi_\caD\left(\sum_i a_i \dd_U b^1_i \cdots \dd_U b^n_i\right) = \sum_i \pi(a_i) [ \caD, \pi(b^1_i)] \cdots [ \caD, \pi(b^n_i)].
\end{equation*}
This suggest to interpret $[\caD, \pi(b)]$ as a differential, but this is impossible: the map $\pi_\caD$ is not a representation of the graded algebra $\Omega^\grast_U(\caA)$, and $\dd_U$ is not represented by the commutator $[\caD, -]$ as a differential. For instance $[\caD, [\caD, \pi(b)]]$ is not necessarily $0$ as required if it were a differential. Using $J$, there is also a representation $\cpi_\caD$ of $\Omega^\grast_U(\caA)$ on the right module structure of $\ehH$: $\Psi \cpi_\caD\left(\sum_i a_i \dd_U b^1_i \cdots \dd_U b^n_i\right) = J \pi_\caD\left(\sum_i a_i \dd_U b^1_i \cdots \dd_U b^n_i \right)^\invast J^{-1} \Psi$.

Let $\hnabla :  \caA \rightarrow \Omega^1_U(\caA)$, with $\omega = \hnabla \bbbone \in \Omega^1_U(\caA)$, be a \nc connection on the right $\caA$-module $\caA$ for the universal differential calculus. Using the bimodule structure on $\ehH$, we have the natural isomorphism of bimodules $\ehH \simeq \caA \otimes_\caA \ehH \otimes_\caA \caA$, where $\Psi \in \ehH$ is identified with $\bbbone \otimes \Psi \otimes \bbbone$. We define the modified Dirac operator $\caD_\omega$ on $\ehH$ by
\begin{equation*}
\caD_\omega (\Psi) = \pi_\caD(\omega) \Psi \otimes \bbbone + \bbbone \otimes \caD \Psi \otimes \bbbone + \epsilon' \bbbone \otimes \Psi \cpi_\caD(\omega)^\invast,
\end{equation*}
for any $\Psi \in \ehH$. This operator can also be written $\caD_\omega = \caD + \pi_\caD(\omega) + \epsilon' J \pi_\caD(\omega) J^{-1}$. 

Let $u \in U(\caA)$. As a gauge transformation (defined as an inner symmetry) of the spectral triple, it acts on $\caD_\omega$ as
\begin{equation*}
(\caD_\omega)^u = \underbrace{\caD + \pi(u) [\caD, \pi(u)^\invast] + \epsilon' J \pi(u) [\caD, \pi(u)^\invast] J^{-1}}_{\caD^u}
+
\pi(u) \pi_\caD(\omega) \pi(u)^\invast + \epsilon' J \pi(u) \pi_\caD(\omega) \pi(u)^\invast J^{-1}.
\end{equation*}
But $u$ is also a gauge transformation as an automorphism of the right module $\caA$, $a \mapsto ua$. The gauge transformation of the connection $1$-form $\omega$ is $\omega^u = u \omega u^\invast + u \dd_U u^\invast$, and the associated modified Dirac operator $\caD_{\omega^u}$ on $\ehH$ is then given  by
\begin{equation*}
\caD_{\omega^u} = \caD + \pi_\caD( u \omega u^\invast + u \dd_U u^\invast) + \epsilon' J \pi_\caD( u \omega u^\invast + u \dd_U u^\invast) J^{-1}.
\end{equation*}
Developing this relation shows that it is $(\caD_\omega)^u$. This means that the two implementations of gauge transformations coincide.

It can be shown that $(\algA, \ehH, \caD_\omega)$ is a spectral triple. The replacement of $\caD$ by $\caD_\omega$ is called an inner fluctuation in the space of Dirac operators associated to the couple $(\algA, \ehH)$. In this approach, gauge fields are inner fluctuations in the space of Dirac operators. Notice that the original Dirac operator $\caD$  is (in general) an unbounded operator on $\ehH$, while inner fluctuations $\pi_\caD(\omega)$ are bounded operators by hypothesis. This implies in particular that the $K$-homology class defined by $\caD$ and $\caD_\omega$ are the same, since inner fluctuations then reduce to compact perturbations of the Fredholm operator associated to $\caD$.

Let consider the case of a spectral triple associated to a spin geometry, where locally $\caD = i\gamma^\mu \partial_\mu$. Then an inner fluctuation corresponds to the twist of the Dirac operator by a connection defined on a vector bundle $\varE$. This procedure consists to replace $\slashed{S}$ by $\slashed{S} \otimes \varE$ and to define $\caD_A = i\gamma^\mu (\partial_\mu + A_\mu)$ on this space using a connection $A$ on $\varE$. This is the minimal replacement $\partial_\mu \mapsto D_\mu$ explained in \ref{sec Ehresmann connections}.

The spectral properties of the Dirac operator $\caD_\omega$ is used to define a gauge invariant action functional $\Act[\caD_\omega]$ using the spectral action principle \cite{ChamConn97a}:
\begin{equation*}
\Act[\caD_\omega] = \tr \chi( \caD_\omega^2/\Lambda),
\end{equation*}
where $\tr$ is the trace on operators on $\ehH$, $\chi$ is a positive and even smooth function $\gR \rightarrow \gR$, and $\Lambda$ is a real (energy) cutoff which helps to make this trace well-behaved. For asymptotically large $\Lambda$, this action can be evaluated using heat kernel expansion. The action functional $\Act[\caD_\omega]$ produces the dynamical part for the gauge fields of the theory, and one has to add the minimal coupling with fermions in the form $\langle \Psi, \caD_\omega \Psi \rangle$ for $\Psi \in \ehH$ to get a complete functional action.

\smallskip
This procedure has been applied to propose a \nc Standard Model of particle physics, which gives a clear geometric origin for the scalar fields used in the SSBM \cite{ChamConnMarc07a, Conn07a, ChamConn08a, ChamConn12a}. This model relies on a so-called ``almost commutative geometry'', which consists to use an algebra of the type $\caA = C^\infty(\varM) \otimes \caA_F$ for a spin manifold $\varM$ and finite dimensional algebra $\caA_F$, for instance a sum of matrix algebras. The total spectral triple $(\algA, \ehH, \caD)$ is the product of a commutative spectral triple $(C(\varM), L^2(\slashed{S}), \slashed{\partial})$ with a ``finite spectral triple'' $(\algA_F, \ehH_F, \caD_F)$:
\begin{align*}
\algA &= C(\varM) \otimes \algA_F,
&
\ehH &= L^2(\slashed{S}) \otimes \ehH_F,
&
\caD &= \slashed{\partial} \otimes 1 + \gamma_\varM \otimes \caD_F,
&
\gamma &= \gamma_\varM \otimes \gamma_F,
&
J &= J_\varM \otimes J_F.
\end{align*}
For the Standard Model, one takes $\caA_F = \gC \oplus \gH \oplus M_3(\gC)$ and $\ehH_F = M_4(\gC) \oplus M_4(\gC) \simeq \gC^{32}$, which contains exactly all the fields of a family of fermions in a see-saw model. At the end, the full Hilbert space is taken to be $\ehH_F^{3}$ to account for the $3$ families of particles. Using this geometry, inner fluctuations can be described, and the most general Dirac operator is
\begin{equation*}
\caD_\omega = \slashed{\partial} + i \gamma^\mu A_\mu + \gamma^5 \caD_F + \gamma^5 \Phi,
\end{equation*}
where the $A_\mu$'s contain all the $U(1) \times SU(2) \times SU(3)$ gauge fields, and $\Phi$ is a doublet of scalar fields, which enters into the SSBM. Note that this (re)formulation of the Standard Model is more constrained than the original one (see for instance \cite{JureKrajSchuStep07a}).

This model describes in the same Lagrangian the Standard Model of particle physics, and the Einstein's theory of gravitation. Indeed, the group of symmetries on which this Lagrangian is invariant is $\Aut(\caA)$, which fits in the short exact sequence of groups
\begin{equation}
\label{eq-ses-automorphism-groups}
\xymatrix@R=0pt@C=15pt{ 
{\algun} \ar[r] & {\Inn(\caA)} \ar[r] & {\Aut(\caA)} \ar[r] & {\Out(\caA)} \ar[r] & {\algun}
 }
\end{equation}
where $\Inn(\caA)$ are inner automorphisms, the gauge transformations, and $\Out(\caA)$ are outer automorphisms, the diffeomorphisms of $\varM$.

\subsection{Derivation-based \ncg}
\label{sec Derivation-based NCG}

Derivation-based \nc geometry was defined in \cite{Dubo88a}, and it has been studied for various algebras, for instance in \cite{DuboKernMado90a, DuboKernMado90b, Mass96a, DuboMass98a, Mass99a, DuboMich94a, DuboMich96a, DuboMich97a, CagnMassWall11a}. See also \cite{Dubo01a, Mass08c, Mass08b} for reviews. The idea is to introduce a natural differential calculus which is based on the derivations of the associative algebra.

Let $\algA$ be an associative algebra with unit $\bbbone$, and let $\caZ(\algA) = \{ a \in \algA \ / \ ab = ba, \forall b \in \algA \}$ its center. The space of derivations of $\algA$ is
\begin{equation*}
\der(\algA) = \{ \kX : \algA \rightarrow \algA \ / \ \kX \text{ linear}, \kX\cdotaction(ab) = (\kX\cdotaction a) b + a (\kX\cdotaction b), \forall a,b\in \algA\}.
\end{equation*}
This vector space is a Lie algebra for the bracket $[\kX, \kY ]a = \kX  \kY a - \kY \kX a$ for all $\kX,\kY \in \der(\algA)$, and a $\caZ(\algA)$-module for the product $(f\kX )\cdotaction a = f(\kX\cdotaction a)$ for all $f \in \caZ(\algA)$ and $\kX \in \der(\algA)$. The subspace 
\begin{equation*}
\Int(\algA) = \{ \ad_a : b \mapsto [a,b]\ / \ a \in \algA\} \subset \der(\algA)
\end{equation*}
is called the vector space of inner derivations: it is a Lie ideal and a $\caZ(\algA)$-submodule. The quotient $\Out(\algA)=\der(\algA)/\Int(\algA)$ gives rise to the short exact sequence of Lie algebras and $\caZ(\algA)$-modules
\begin{equation}
\label{eq-sesderivations}
\xymatrix@1@C=15pt{{\algzero} \ar[r] & {\Int(\algA)} \ar[r] & {\der(\algA)} \ar[r] & {\Out(\algA)} \ar[r] & {\algzero}}.
\end{equation}
$\Out(\algA)$ is called the space of outer derivations of $\algA$. This short exact sequence is the infinitesimal version of \eqref{eq-ses-automorphism-groups}. If $\algA$ is commutative, there are no inner derivations, and the space of outer derivations is the space of all derivations.

In case $\algA$ has an involution, a derivation $\kX \in \der(\algA)$ is said to be real when $(\kX a)^\invast = \kX a^\invast$ for any $a\in \algA$, and we denote by $\der_\gR(\algA)$ the space of real derivations.

Let $\underline{\Omega}^n_\der(\algA)$ be the vector space of $\caZ(\algA)$-multilinear antisymmetric maps from $\der(\algA)^n$ to $\algA$, with $\underline{\Omega}^0_\der(\algA) = \algA$. Then the total space
\begin{equation*}
\underline{\Omega}^\grast_\der(\algA) =\bigoplus_{n \geq 0} \underline{\Omega}^n_\der(\algA)
\end{equation*}
gets a structure of $\gN$-graded differential algebra for the product
\begin{equation*}
(\omega\eta)(\kX_1, \dots, \kX_{p+q}) =
 \frac{1}{p!q!} \sum_{\sigma\in \kS_{p+q}} (-1)^{\sign(\sigma)} \omega(\kX_{\sigma(1)}, \dots, \kX_{\sigma(p)}) \eta(\kX_{\sigma(p+1)}, \dots, \kX_{\sigma(p+q)})
\end{equation*}
for any $\kX_i \in \der(\algA)$. A differential $\hd$ is defined by the so-called Koszul formula
\begin{multline*}
\hd\omega(\kX_1, \dots , \kX_{n+1}) = \sum_{i=1}^{n+1} (-1)^{i+1} \kX_i\cdotaction \omega( \kX_1, \dots \omi{i} \dots, \kX_{n+1}) \\[-5pt]
 + \sum_{1\leq i < j \leq n+1} (-1)^{i+j} \omega( [\kX_i, \kX_j], \dots \omi{i} \dots \omi{j} \dots , \kX_{n+1}). 
\end{multline*}
We denote by $\Omega^\grast_\der(\algA)$ the graded differential sub-algebra of $\underline{\Omega}^\grast_\der(\algA)$ generated in degree $0$ by $\algA$. Every element in $\Omega^n_\der(\algA)$ is a sum of terms of the form $a_0 \hd a_1 \cdots \hd a_n$ for $a_0, \dots, a_n \in \algA$. We will refer to $\underline{\Omega}^\grast_\der(\algA)$ as the maximal differential calculus and to $\Omega^\grast_\der(\algA)$ as the minimal one. The minimal differential calculus being generated by $\algA$, it is a quotient of the universal differential calculus, but the maximal differential calculus can contain elements which are not in this quotient.

The previous construction is motivated by the ``commutative'' situation: let $\algA = C^\infty(\varM)$ for a smooth compact manifold $\varM$, then $\caZ(\algA) = C^\infty(\varM)$; $\der(\algA) = \Gamma(T\varM)$ is the Lie algebra of vector fields on $\varM$; $\Int(\algA) = \algzero$; $\Out(\algA) = \Gamma(T\varM)$; and $\Omega^\grast_\der(\algA) = \underline{\Omega}^\grast_\der(\algA) = \Omega^\grast(\varM)$ is the graded differential algebra of de~Rham forms on $\varM$.

\NC connections constructed in this framework look very much like ordinary connections. Let $\modM$ be a right $\algA$-module. Then a \nc connection on $\modM$ is a linear map $\hnabla_\kX : \modM \rightarrow \modM$ defined for any $\kX \in \der(\algA)$, such that for all $\kX,\kY \in \der(\algA)$, $a \in \algA$, $m \in \modM$, and $f \in \caZ(\algA)$ one has (compare with \eqref{eq-definition-rules-nabla}):
\begin{align*}
\hnabla_\kX (m a) &=  (\hnabla_\kX m) a + m(\kX\cdotaction a),
&
\hnabla_{f\kX} m &= f \hnabla_\kX m,
&
\hnabla_{\kX + \kY} m &= \hnabla_\kX m + \hnabla_\kY m.
\end{align*}
Its curvature is the right $\algA$-module morphism $\hR(\kX, \kY) : \modM \rightarrow \modM$ defined for any $\kX, \kY \in \der(\algA)$ by $\hR(\kX, \kY) m = [ \hnabla_\kX, \hnabla_\kY ] m - \hnabla_{[\kX, \kY]}m$, which is the obstruction on $\hnabla$ to be a morphism of Lie algebras between $\der(\algA)$ and the space of (differential) operators on $\modM$. For the module $\modM = \algA$, $\kX \mapsto \hnabla_\kX \bbbone = \homega(\kX) \in \algA$ defines the connection $1$-form $\homega$ of $\hnabla$.

\medskip
Let us now consider the finite dimensional algebra $\algA = M_n(\gC) = M_n$ of $n\times n$ complex matrices as a particular example. Its derivation-based differential calculus can be described in details (see \cite{Dubo88a, DuboKernMado90a, Mass08c}). From well-known results in algebra, one has: $\caZ(M_n) = \gC$, and $\der(M_n) = \Int(M_n) \simeq \ksl_n =\ksl_n(\gC)$, where $\ksl_n(\gC)$ is the $n^2-1$-dimensional Lie algebra of traceless complex $n \times n$ matrices. The isomorphism associates to any $\gamma \in \ksl_n$ the derivation $\ad_\gamma : a \mapsto [\gamma, a]$. Since $\der(\algA) = \Int(\algA)$, one has $\Out(M_n) = \algzero$: this is the opposite situation to the one encountered for commutative algebras. Adjointness defines an involution, and the space of real derivations is $\der_\gR(M_n) = \ksu(n)$, the Lie algebra of traceless Hermitian matrices, where the identification is given by $\gamma \mapsto \ad_{i \gamma}$ for any $\gamma \in \ksu(n)$. The associated derivation-based differential calculus can be described as
\begin{equation*}
\underline{\Omega}^\grast_\der(M_n) = \Omega^\grast_\der(M_n) \simeq M_n \otimes \exter^\grast \ksl_n^\dualast,
\end{equation*}
and its differential, denoted by $\dd'$, identifies with the differential of the Chevalley-Eilenberg complex of the Lie algebra $\ksl_n$ represented on $M_n$ by the adjoint representation (commutator) \cite{CartEile56a, Weib97a, Mass08a}. Since the maximal and minimal differential calculi coincide, we will use the notation $\Omega^\grast_\der(M_n)$ for this differential calculus.

There is a canonical \nc $1$-form $i\theta \in \Omega^1_\der(M_n)$ defined, for any $\gamma \in M_n(\gC)$, by
\begin{equation*}
i\theta(\ad_{\gamma}) = \textstyle\gamma - \frac{1}{n} \tr (\gamma)\bbbone,
\end{equation*}
which plays an important role. First, it makes explicit the isomorphism $\Int(M_n) \xrightarrow{\simeq} \ksl_n$. Moreover, it satisfies $\dd' a = [i\theta, a] \in \Omega^1_\der(M_n)$ for any $a \in M_n$ (this relation is no more true in higher degrees). The relation $\dd' (i\theta) - (i\theta)^2 = 0$ makes $i\theta$ looks very much like the Maurer-Cartan form in the geometry of Lie groups $SL_n(\gC)$.

Denote by $\{E_k\}_{k=1, \dots, n^2-1}$ a basis of $\ksl_n$ of traceless Hermitian matrices, and let $C^m_{k \ell}$ be the real structure constants of $\ksl_n$ in this basis: $[E_k, E_\ell] = -i C^m_{k \ell} E_m$. We can adjoin the unit $\bbbone$ to the $E_k$'s to get a basis for $M_n$. The $n^2-1$ (real) derivations $\partial_k = \ad_{i E_k}$ define a basis of $\der(M_n) \simeq \ksl_n$, and one has $[\partial_k, \partial_\ell] = C^m_{k \ell} \partial_m$. Let $\{\theta^\ell\}$ be the dual basis in $\ksl_n^\dualast$: $\theta^\ell(\partial_k) = \delta^\ell_k$. It generates a basis for the exterior algebra $\exter^\grast \ksl_n^\dualast$, where by definition one has $\theta^\ell \theta^k = - \theta^k \theta^\ell$.

Any \nc $p$-form decomposes as a finite sum of terms of the form $a \otimes \theta^{k_1} \cdots \theta^{k_p}$ for $k_1 < \cdots < k_p$ and $a = a^k E_k + a^0 \bbbone \in M_n$, for instance one has $i\theta = i E_k \otimes \theta^k \in M_n \otimes \exter^1 \ksl_n^\dualast$. The differential $\dd'$ is given on the generators of $\Omega^\grast_\der(M_n)$ by $\dd' \bbbone = 0$, $\dd' E_k = -C^m_{k\ell} E_m \otimes \theta^\ell$, and $\dd' \theta^k = -\tfrac{1}{2} C^k_{\ell m} \theta^\ell \theta^m$.

Let us consider a gauge theory for the $\algA$-module $\modM = \algA$ equipped with the Hermitian structure $\langle a, b \rangle = a^\invast b$. Then, one can show that the canonical $1$-form $i\theta$ defines a non trivial canonical \nc connection given by $\hnabla^{-i\theta}_\kX a = - a i\theta(\kX) = \kX\cdotaction a - i\theta(\kX) a = -a \gamma$ for any $a \in \algA$ and any $\kX = \ad_\gamma \in \der(M_n)$ (with $\tr \gamma = 0$). This Hermitian connection is special in the sense that it is gauge invariant (for the action of $g \in U(\algA) = U(n)$, the group of unitary matrices), its curvature is zero, but it is not pure gauge. $\hnabla^{-i\theta}$ defines a particular and preferred element in the affine space of \nc connections, and we can decompose any \nc connection as
\begin{equation*}
\hnabla_\kX a = \hnabla^{-i\theta}_\kX a + A(\kX) a = (A - i\theta)(\kX) a,
\end{equation*}
for a \nc $1$-form $A = A_k \otimes \theta^k \in \Omega^1_\der(M_n)$. Such a connection is Hermitian if and only if the $A_k$'s are anti-Hermitian matrices. Under a gauge transformation $g \in U(n)$, one has $A_k \mapsto g^{-1} A_k g$: the inhomogeneous term has been absorbed by $-i\theta$. Then the curvature of $\hnabla$ is the multiplication on the left by the $2$-form
\begin{equation*}
F = \tfrac{1}{2}( [A_k, A_\ell] - C^m_{k \ell} A_m) \otimes \theta^k \theta^\ell,
\end{equation*}
and the matrices $F_{k \ell} = [A_k, A_\ell] - C^m_{k \ell} A_m$ are anti-Hermitian. A natural action functional for this connection is 
\begin{equation*}
\Act[A] = - \tfrac{1}{8n}  \tr \left( F_{k \ell}F^{k \ell}\right).
\end{equation*}
One has $\Act[A] \geq 0$ and its minimum, which is $0$, is obtained in two situations: $\hnabla$ is a pure gauge connection or $\hnabla = \hnabla^{-i\theta}$. For more details, we refer to \cite{DuboKernMado90a}.

\medskip
Let us now briefly describe the situation for the algebra $\algA = C^\infty(\varM) \otimes M_n(\gC)$ where $\varM$ is a $m$-dimensional compact smooth manifold. This \nc geometry was first considered in \cite{DuboKernMado90b}, to which we refer for further details. The main results are: 
\begin{itemize}
\item $\caZ(\algA) = C^\infty(\varM)$, where $f \in C^\infty(\varM)$ identifies with $f \bbbone_n$ ($\bbbone_n$ is the identity matrix in $M_n$).

\item $\der(\algA) = [\der(C^\infty(\varM))\otimes \bbbone_n ] \oplus [ C^\infty(\varM) \otimes \der(M_n) ] = \Gamma(T\varM) \oplus [C^\infty(\varM) \otimes \ksl_n]$ is a splitting as Lie algebras and $C^\infty(\varM)$-modules. 

\item $\Int(\algA) = \algA_0 = C^\infty(\varM) \otimes \ksl_n$ is the Lie algebra of traceless elements in $\algA$ for the commutator.

\item $\Out(\algA) = \Gamma(T\varM)$.

\item The maximal and minimal differential calculi coincide:
\begin{equation*}
\underline{\Omega}^\grast_\der(\algA) = \Omega^\grast_\der(\algA) = \Omega^\grast(\varM) \otimes \Omega^\grast_\der(M_n),
\end{equation*}
where $\Omega^\grast(\varM)$ is the de~Rham differential calculus on $\varM$. The differential is $\hd = \dd + \dd'$, where $\dd$ is the de~Rham differential and $\dd'$ is the differential introduced in the previous example.
\end{itemize}
Let us use the notation $\der(\algA) \ni \kX = X \oplus \gamma$ for $X \in \Gamma(T\varM)$ and $\gamma : \varM \rightarrow\ksl_n$. Then the \nc $1$-form $i\theta$ defined by $i\theta(X \oplus \gamma) = \gamma$ gives the splitting of \eqref{eq-sesderivations} as Lie algebras and $C^\infty(\varM)$-modules:
\begin{equation}
\label{eq-splittingsecderivationstrivialcase}
\xymatrix@1@C=25pt{{\algzero} \ar[r] & {\algA_0} \ar[r] & {\der(\algA)} \ar[r] \ar@/_0.7pc/[l]_-{i\theta}& {\Gamma(T\varM)} \ar[r] & {\algzero}}.
\end{equation}

Let us describe gauge theories for the right $\algA$-module $\modM = \algA$ equipped with the Hermitian structure $\langle a, b \rangle = a^\invast b$. Let $\hnabla^{-i\theta}$ be the canonical connection defined by $\hnabla^{-i\theta}_\kX a = X \cdotaction a - a \gamma$ for any $\kX = X \oplus \gamma \in \der(\algA)$. Its curvature is zero, but contrary to the previous example, it is not gauge invariant. We can use it as a particular point in the affine space of \nc connections.

Any \nc connection $\hnabla$ on $\algA$ can be written as $\hnabla_\kX a = \hnabla^{-i\theta}_\kX a + A(\kX) a$ with $A \in \Omega^1_\der(\algA)$. Let us decompose $A$ as 
\begin{equation}
\label{eq-split-ab}
A(X \oplus \gamma) = \ka(X) + \kb(\gamma)
\end{equation}
for $\ka = \ka_\mu \dd x^\mu \in M_n \otimes \Omega^1(\varM)$ and $\kb = \kb_k \theta^k \in C^\infty(\varM) \otimes M_n \otimes \exter^1 \ksl_n^\dualast$. $\hnabla$ is Hermitian when $A$ takes its values in anti-Hermitian matrices, and under a gauge transformation $g : \varM \rightarrow U(n)$, one has
\begin{align*}
\ka_\mu &\mapsto g^{-1} \ka_\mu g + g^{-1} \partial_\mu g,
&
\kb_k &\mapsto g^{-1} \kb g.
\end{align*}
The curvature of $\hnabla$ is the \nc $2$-form
\begin{equation*}
F = \textstyle\frac{1}{2}(\partial_\mu \ka_\nu - \partial_\nu \ka_\mu + [ \ka_\mu, \ka_\nu])\dd x^\mu \dd x^\nu + (\partial_\mu \kb_k + [ \ka_\mu, \kb_k]) \dd x^\mu \theta^k + \frac{1}{2}([\kb_k, \kb_\ell] - C^m_{k \ell} \kb_m) \theta^k \theta^\ell.
\end{equation*}
One can define the following natural gauge invariant action functional for $\hnabla$:
\begin{multline}
\label{eq-ncg-act}
\Act[A] = -\frac{1}{4n} \int \dd x \tr\bigg\{ \sum_{\mu,\nu}(\partial_\mu \ka_\nu - \partial_\nu \ka_\mu + [ \ka_\mu, \ka_\nu])^2 \\[-5pt]
-\frac{\mu^2}{2n} \sum_{\mu,k} (\partial_\mu \kb_k + [ \ka_\mu, \kb_k])^2 
-\frac{\mu^4}{4n} \sum_{k, \ell} ([\kb_k, \kb_\ell] - C^m_{k \ell} \kb_m)^2 \bigg\}.
\end{multline}
where $\mu$ is a positive constant which, in physical natural units, has the dimension of a mass. 

The integrand of $\Act[A]$ can be zero on two gauge orbits:
\begin{enumerate}
\item $\ka = g^{-1} \dd g$ and $\kb_k=0$ is the gauge orbit of $\hnabla = \hnabla^{-i\theta}$.

\item $\ka = g^{-1} \dd g$ and $\kb_k= i g^{-1} E_k g$ is the gauge orbit of $\hnabla_\kX a = \kX\cdotaction a$.
\end{enumerate}
 The configurations with arbitrary $\ka$ and $\kb_k = i E_k$ describe connections where the fields $a_\mu$ have mass terms coming from the second term in the Lagrangian. This reveals a SSBM where the $\kb_k$'s plays the role of scalar fields coupled to the $U(n)$-Yang-Mills fields $\ka_\mu$ through a covariant derivative in the adjoint representation, and where the last term in the integrand is a quadratic potential which admits minima for non zero configurations. These scalar fields are not introduced by hand in the Lagrangian since they are part of the \nc connection along the purely algebraic directions. $\Act[A]$ describes a Yang-Mills-Higgs model. See \cite{DuboKernMado90b} for more details.

\medskip
There is a natural generalization of this example in the following form. Let $\varP$ be a $SU(n)$-principal fiber bundle over $\varM$ (as before), and let $\varE$ be the associated vector bundle for the fundamental representation of $SU(n)$ on $\gC^n$. Consider $\algA$ as the associative algebra of smooth sections of the vector bundle $\End(\varE) = \varE \otimes \varE^\dualast$, whose fiber is $M_n(\gC)$.  This is the algebra of endomorphisms of $\varE$. Its \nc geometry has been studied in \cite{DuboMass98a, Mass99a, MassSeri05a}, see \cite{Mass08c} for a review. 

When $\varP = \varM \times SU(n)$ is the trivial bundle, then $\algA = C^\infty(\varM) \otimes M_n(\gC)$ as before. When $\varP$ has a non trivial topology, one can always identify locally $\algA$ as $C^\infty(\varU) \otimes M_n(\gC)$ for an open subset $\varU \subset \varM$, so that the results of the previous example are still useful. Fiberwise, one can define an involution, the trace map $\tr : \algA \to C^\infty(\varM)$, and a determinant $\det : \algA \to C^\infty(\varM)$. 

The main results on this \nc geometry are:
\begin{itemize}
\item $\caZ(\algA) = C^\infty(\varM)$.

\item The projection $\rho : \der(\algA) \rightarrow \der(\algA)/\Int(\algA) = \Out(\algA)$  is the restriction of derivations $\kX \in \der(\algA)$ to $\caZ(\algA) = C^\infty(\varM)$. We will use the typographic convention $\rho(\kX) = X$.

\item $\Out(\algA) \simeq \der(C^\infty(\varM)) = \Gamma(T\varM)$.

\item $\Int(\algA)$ is isomorphic to $\algA_0$, the traceless elements in $\algA$.

\item The short exact sequence \eqref{eq-sesderivations} looks like
\begin{equation}
\label{eq-ses-endo}
\xymatrix@1{ {\algzero} \ar[r] & {\algA_0} \ar[r]^-{\ad} & {\der(\algA)} \ar[r]^-{\rho} & {\Gamma(T\varM)} \ar[r] & {\algzero} }.
\end{equation}

\item $\underline{\Omega}^\grast_\der(\algA) = \Omega^\grast_\der(\algA)$. We denote by $\hd$ its differential.

\item There is a well defined map of $C^\infty(\varM)$-modules $i\theta : \Int(\algA) \rightarrow \algA_0$ given by $\ad_\gamma \mapsto \textstyle\gamma - \frac{1}{n}\tr(\gamma) \bbbone$.

\end{itemize}

The map $i\theta$ does not extend to $\der(\algA)$. It is possible to defined a splitting of \eqref{eq-ses-endo} by the following procedure. Let $\nabla^\varE$ be any (ordinary) $SU(n)$-connection on $\varE$, and let $\nabla$ be its associated connection on $\End(\varE)$. Then for any $X \in \Gamma(T\varM)$, $\nabla_X \in \der(\algA)$, and the map $X \mapsto \nabla_X$ is a splitting of \eqref{eq-ses-endo} as $C^\infty(\varM)$-modules, but not as Lie algebras, since the obstruction $R(X,Y) = [\nabla_X, \nabla_Y] - \nabla_{[X,Y]}$ is precisely the curvature of $\nabla$. For any $\kX \in \der(\algA)$, let $X = \rho(\kX)$. Then $\rho(\kX - \nabla_X) = 0$, so that there is a $\alpha(\kX) \in \algA_0$ such that $\kX = \nabla_X - \ad_{\alpha(\kX)}$. The map $\kX \mapsto \alpha(\kX)$ belongs to $\Omega^1_\der(\algA)$ and satisfies the normalization $\alpha(\ad_\gamma) = -\gamma$ for any $\gamma \in \algA_0$. 

This map realizes an isomorphism between the space of $SU(n)$-connections $\nabla^\varE$ on $\varE$ and the space of traceless anti-Hermitian \nc $1$-forms $\alpha$ on $\algA$ such that $\alpha(\ad_\gamma) = -\gamma$. The \nc $1$-form $\alpha$ is defined globally on $\varM$, and it completes (in a new space of forms) the last description proposed in \ref{sec Ehresmann connections} on connections and curvatures in ordinary differential geometry. It can be shown that $\alpha$ is defined in terms of the local  trivializations $A_i$ of the connection $1$-form associated to $\nabla^\varE$ (see \cite{Mass99a, Mass08c}). In the same way, the \nc $2$-form $\Omega(\kX, \kY) = \hd\alpha(\kX, \kY) + [\alpha(\kX), \alpha(\kY) ]$ depends only on the projections $X$ and $Y$ of $\kX$ and $\kY$: as a section of $\exter^2 T^\ast \varM \otimes \End(\varE)$, it identifies with $\gF$, \textsl{i.e.} the curvature $R^\varE$ of $\nabla^\varE$.

Notice that the gauge group $\caG(\varP)$ of $\varP$ is precisely $SU(\algA) \subset \algA$, the unitary elements in $\algA$ with determinant $1$. The action of $u \in \caG(\varP) = SU(\algA)$ on $\nabla^\varE$ induces the action $\alpha \mapsto \alpha^u = u^{-1} \alpha u + u^{-1} \hd u$ on $\alpha$.

Let us now consider \nc connections on the $\algA$-module $\modM = \algA$ equipped with the Hermitian structure $\langle a, b \rangle = a^\invast b$. From the general theory we know that any \nc $1$-form $\homega$ defines a \nc connection by $\hnabla_\kX a = \kX\cdotaction a + \homega(\kX) a$ for any $a \in \modM = \algA$ and $\kX \in \der(\algA)$. In particular, the \nc $1$-form $\alpha$ associated to $\nabla^\varE$ defines a \nc connection $\hnabla^\alpha$ which can be written as $\hnabla^\alpha_\kX a = \nabla_X a + a \alpha(\kX)$. Then $\hnabla^\alpha$ is compatible with the Hermitian structure, its curvature is $\hR^\alpha(\kX, \kY) = R^\varE(X,Y) = \hd\alpha(\kX, \kY) + [\alpha(\kX), \alpha(\kY) ]$, and a $SU(\algA)$-\nc gauge transformation on $\hnabla^\alpha$ is exactly a (ordinary) gauge transformation on $\nabla^\varE$ (here we use the fact that the two gauge groups are the same). The main result of this construction is that the space of \nc connections on the right module $\algA$ compatible with the Hermitian structure $(a,b) \mapsto a^\ast b$ contains the space of ordinary $SU(n)$-connections on $\varE$, and this inclusion is compatible with the corresponding definitions of curvature and $SU(\algA)=\caG(\varP)$ gauge transformations.

A \nc connection $\homega$ describes an ordinary connection if and only if it is normalized on inner derivations: $\homega(\ad_\gamma) = -\gamma$. This implies that \nc connections have more degrees of freedom that ordinary connections. In gauge field models, these degrees of freedom describe scalar fields which induce (as in the case $\algA = C^\infty(\varM) \otimes M_n(\gC)$) a SSBM. We refer to \cite{Mass08c} for more details.

\section{Transitive Lie algebroids}
\label{sec Transitive Lie algebroids}

Lie algebroids have been defined and studied in relation with classical mechanics and its various modern mathematical formulations, like Poisson geometry and symplectic manifolds (see \cite{Kosm80a, AlmeMoli85a, Kara86a, Wein87a}, \cite{Marl02a, Kosm08a} and references in \cite{Mack05a, CraiFern06a}). This approach considers a Lie algebroid as a generalization of the tangent bundle, on which a Lie bracket is defined. Our approach departs from this geometrical point of view. We would like to consider a Lie algebroid (more precisely a transitive Lie algebroid) as an algebraic replacement for a principal vector bundle, from which it is possible to construct gauge field theories. This program has been proposed in \cite{LazzMass12a} where the useful notion of connection has been studied, and it has been pursued in \cite{FourLazzMass13a}, where the necessary tools to build gauge fields theories have been defined.

\subsection{Generalities on transitive Lie algebroids}
\label{sec Generalities on transitive Lie algebroids}

The usual definition of Lie algebroids consists in the following geometrical description. A Lie algebroid $(\varA, \rho)$ is a vector bundle $\varA$ over a smooth $m$-dimensional manifold $\varM$ equipped with two structures:
\begin{enumerate}
\item a structure of Lie algebra on the space of smooth sections $\Gamma(\varA)$,
\item a vector bundle morphism $\rho : \varA \to T\varM$, called the anchor, such that
\begin{align*}
\rho([\kX, \kY]) &= [\rho(\kX), \rho(\kY)],
&
[\kX, f \kY] &= f[\kX, \kY] + (\rho(\kX)\cdotaction f)\, \kY,
\end{align*}
for any $\kX, \kY \in \Gamma(\varA)$ and $f \in C^\infty(\varM)$.
\end{enumerate}
 
The following (equivalent) algebraic definition of Lie algebroids will be used in the following. The geometric structure is ignored in favor of the algebraic structure, as in \nc geometry, from which we will borrow some ideas and constructions in the following. 

A Lie algebroid $\lieA$ is a finite projective module over $C^\infty(\varM)$ equipped with a Lie bracket $[-,-]$ and a $C^\infty(\varM)$-linear Lie morphism, the anchor $\rho : \lieA \rightarrow \Gamma(T\varM)$, such that $[\kX, f \kY] = f [\kX, \kY] + (\rho(\kX)\cdotaction f) \kY$ for any $\kX, \kY \in \lieA$ and $f \in C^\infty(\varM)$. 

A Lie algebroid $\lieA \xrightarrow{\rho} \Gamma(T\varM)$ is transitive if $\rho$ is surjective. The kernel $\lieL = \ker \rho$ of a transitive Lie algebroid is itself a Lie algebroid with null anchor. Moreover, there exists a locally trivial bundle in Lie algebras $\varL$ such that $\lieL = \Gamma(\varL)$. A transitive Lie algebroid defines a short exact sequence of Lie algebras and $C^\infty(\varM)$-modules
\begin{equation}
\label{eq-sectransitiveliealgebroid}
\xymatrix@1{{\algzero} \ar[r] & {\lieL} \ar[r]^-{\iota} & {\lieA} \ar[r]^-{\rho} & {\Gamma(T\varM)} \ar[r] & {\algzero}}.
\end{equation}
From a gauge field theory point of view, this short exact sequence must be looked at as an infinitesimal version of the sequence \eqref{eq-principal-bundle} defining a principal fiber bundle.

The kernel $\lieL$ will be referred to as the ``inner'' part of $\lieA$. This terminology is inspired by the physical applications we have in mind, where $\Gamma(T\varM)$ will refer to (infinitesimal) symmetries on space-time (``outer'' symmetries) and $\lieL$ to (infinitesimal) inner symmetries \textsl{i.e.} infinitesimal gauge symmetries. Compare this with the physical interpretation of the short exact sequences \eqref{eq-ses-automorphism-groups} and \eqref{eq-ses-endo}.

A morphism between two Lie algebroids $(\lieA, \rho_\lieA)$ and $(\lieB, \rho_\lieB)$ is a morphism of Lie algebras and $C^\infty(\varM)$-modules $\varphi: \lieA \rightarrow \lieB$ compatible with the anchors: $\rho_\lieB \circ \varphi = \rho_\lieA$. 

The following example of transitive Lie algebroid is fundamental to define the correct notion of representation. Let $\varE$ be a vector bundle over $\varM$, and let $\algA(\varE)$ be the associative algebra of endomorphisms of $\varE$ as in the end of  \ref{sec Derivation-based NCG}. Denote by $\kD(\varE)$ the space of first-order differential operators on $\varE$ with scalar symbols. Then the restricted symbol map $\sigma : \kD(\varE) \rightarrow \Gamma(T\varM)$ produces the short exact sequence 
\begin{equation*}
\xymatrix@1{{\algzero} \ar[r] & {\algA(\varE)} \ar[r]^-{\iota} & {\kD(\varE)} \ar[r]^-{\sigma} & {\Gamma(T\varM)} \ar[r] & {\algzero}}.
\end{equation*}
$\kD(\varE)$ is the transitive Lie algebroid of derivations of $\varE$ \cite{KosmMack02a, Kosm80a}. A representation of a transitive Lie algebroid $\lieA \xrightarrow{\rho} \Gamma(T\varM)$ on a vector bundle $\varE \rightarrow \varM$ is a morphism of Lie algebroids $\phi : \lieA \rightarrow \kD(\varE)$ \cite{Mack05a}. This can be summarized in the commutative diagram of exact rows
\begin{equation}
\label{eq-diagramrepresentation}
\xymatrix
{
{\algzero} \ar[r] & {\lieL} \ar[r]^-{\iota}\ar[d]^-{\phi_\lieL}  & {\lieA} \ar[r]^-{\rho} \ar[d]^-{\phi} & {\Gamma(T\varM)} \ar[r] \ar@{=}[d]  & {\algzero}
\\
{\algzero} \ar[r] & {\algA(\varE)} \ar[r]^-{\iota} & {\kD(\varE)} \ar[r]^-{\sigma} & {\Gamma(T\varM)} \ar[r] & {\algzero}
}
\end{equation}
where $\phi_\lieL : \lieL \rightarrow \algA(\varE)$ is a $C^\infty(\varM)$-linear morphism of Lie algebras. 

The second example permits to embed the ordinary theory of connections on principal fiber bundle into this framework. Let $\varP$ be a $G$-principal fiber bundle over $\varM$ with projection $\pi$. We use the notations of \ref{sec Ehresmann connections}. The two spaces
\begin{align*}
\Gamma_G(T\varP) &= \{ \sfX \in \Gamma(T\varP) \, / \, \tR_{g\,\ast}\sfX = \sfX \text{ for all } g \in G \},
\\
\Gamma_G(\varP, \kg) &= \{ v : P \rightarrow \kg \, / \, v(p \cdotaction g) = \Ad_{g^{-1}} v(p) \text{ for all } g \in G \},
\end{align*}
are naturally Lie algebras and $C^\infty(\varM)$-modules. $\Gamma_G(T\varP)$ is the space of vector fields on $\varP$ which are projectable as vector fields on the base manifold, and $\Gamma_G(\varP, \kg)$ is the space of $(\tR, \Ad)$-equivariant maps $v : P \rightarrow \kg$, which is also the space of sections of the associated vector bundle $\Ad\varP$.

Denote by $\xi^\varP$ the fundamental (vertical) vector field on $\varP$ associated to $\xi \in \kg$. The map $\iota : \Gamma_G(\varP, \kg) \rightarrow \Gamma_G(T\varP)$ defined by
\begin{equation*}
\iota(v)(p) = -v(p)^\varP_{|p} = \left( \frac{d}{dt} p \cdotaction e^{-t v(p)} \right)_{|t=0}
\end{equation*}
is an injective $C^\infty(\varM)$-linear morphism of Lie algebras. The short exact sequence of Lie algebras and $C^\infty(\varM)$-modules
\begin{equation}
\label{eq-ses-atiyah}
\xymatrix@1{{\algzero} \ar[r] & {\Gamma_G(\varP, \kg)} \ar[r]^-{\iota} & {\Gamma_G(T\varP)} \ar[r]^-{\pi_\ast} & {\Gamma(T\varM)} \ar[r] & {\algzero}}
\end{equation}
defines $\Gamma_G(T\varP)$ as a transitive Lie algebroid over $\varM$. This is the Atiyah Lie algebroid associated to $\varP$ \cite{Atiy57a}.

Consider the case where $\varP = \varM \times G$ is trivial. The associated transitive Lie algebroid is denoted by $\tla(\varM, \kg) = \Gamma_G(T\varP)$, and called the Trivial Lie Algebroid on $\varM$ for $\kg$. It is the space of sections of the vector bundle $\varA = T\varM \oplus (\varM \times \kg)$, equipped with the anchor and the bracket
\begin{align*}
\rho(X \oplus \gamma) &= X, 
&
[X \oplus \gamma, Y \oplus \eta] &= [X,Y] \oplus (X \cdotaction \eta - Y \cdotaction \gamma + [\gamma,\eta]),
\end{align*}
for any $X,Y \in \Gamma(T\varM)$ and $\gamma,\eta \in \Gamma(\varM \times \kg) \simeq C^\infty(\varM)\otimes \kg$. The kernel is the space of sections of  the trivial vector bundle $\varL = \varM \times \kg$. The short exact sequence \eqref{eq-sectransitiveliealgebroid} is split as Lie algebras and $C^\infty(\varM)$-modules. The importance of this notion relies on the fact that any transitive Lie algebroid can be described locally as a trivial Lie algebroid $\tla(\varU, \kg)$ over an open subset $\varU \subset \varM$.

\subsection{Differential structures}
\label{sec Differential structures}

Given a representation $\phi : \lieA \to \kD(\varE)$, one can define an associated differential calculus in the following way \cite[Definition~7.1.1]{Mack05a}. For any $p \in \gN$, let $\Omega^p(\lieA, \varE)$ be the linear space of $C^\infty(\varM)$-multilinear antisymmetric maps  $\lieA^p \to \Gamma(\varE)$. For $p=0$ one has $\Omega^0(\lieA, \varE) = \Gamma(\varE)$. The graded space $\Omega^\grast(\lieA, \varE) = \bigoplus_{p \geq 0} \Omega^p(\lieA, \varE)$ is equipped with the natural differential $\hd_\phi : \Omega^p(\lieA, \varE) \rightarrow \Omega^{p+1}(\lieA, \varE)$ defined on $\omega \in \Omega^p(\lieA, \varE)$ by the Koszul formula
\begin{multline*}
(\hd_\phi \omega)(\kX_1, \dots, \kX_{p+1}) = \sum_{i=1}^{p+1} (-1)^{i+1} \phi(\kX_i)\cdotaction\omega(\kX_1, \dots \omi{i} \dots, \kX_{p+1})\\
+ \sum_{1 \leq i < j \leq p+1} (-1)^{i+j} \omega([\kX_i, \kX_j], \kX_1, \dots \omi{i} \dots \omi{j} \dots, \kX_{p+1}).
\end{multline*}
In this definition, $\phi(\kX)\cdotaction \psi$ is the action of the first order differential operator $\phi(\kX)$ on $\psi \in \Gamma(\varE)$. Since $\phi$ is a morphism of Lie algebras, one has $\hd_\phi^2 = 0$. Two particular differential calculi are of interest for the following.

First, let us consider $\varE = \varM \times \gC$, so that $\Gamma(\varE) = C^\infty(\varM)$ (complex valued), $\kD(\varE) = \Gamma(T\varM)$, and $\rho$ defines a natural representation of $\lieA$ on $\varE$. We denote by $(\Omega^\grast(\lieA), \hd_\lieA)$ the associated differential calculus: it is a graded commutative differential algebra. This space of forms is described for instance in  \cite{AriaCrai11a, Crai03a, CraiFern06a, CraiFern09a, Fern03a, Mack05a}. This differential calculus can be defined on any Lie algebroid.

For the second example, $\lieA$ is a transitive Lie algebroid, and we consider $\varE = \varL$, the vector bundle for which $\lieL = \Gamma(\varL)$. There is a natural representation of $\lieA$ on $\varL$, called the adjoint representation, which is defined as follows: for any $\kX \in \lieA$ and any $\ell \in \lieL$, the Lie bracket $\ad_\kX(\ell) = [\kX, \ell]$ is defined to be the unique element in $\lieL$ such that $\iota([\kX, \ell]) = [\kX, \iota(\ell)]$.  We denote by $(\Omega^\grast(\lieA, \lieL), \hd)$ the associated differential calculus. It is a graded differential Lie algebra, and a graded differential module on $\Omega^\grast(\lieA)$.

If $\lieB$ is a Lie algebroid over $\varM$, a Cartan operation of $\lieB$ on $(\Omega^\grast(\lieA, \varE), \hd_\phi)$ is given by the following data \cite{Ginz99a}: for any $\kX \in \lieB$, for any $p \geq 1$, there is a map $i_\kX : \Omega^p(\lieA, \varE) \rightarrow \Omega^{p-1}(\lieA, \varE)$ such that the relations 
\begin{align}
\label{eqs-cartanoperation}
i_{f \kX} &= f i_{\kX},
&
i_\kX i_\kY + i_\kY i_\kX &= 0,
&
[L_\kX, i_\kY] &= i_{[\kX, \kY]},
&
[L_\kX, L_\kY] &= L_{[\kX, \kY]},
\end{align}
hold for any $\kX, \kY \in \lieB$, $f \in C^\infty(\varM)$, where $L_\kX = \hd_\phi i_\kX + i_\kX \hd_\phi$. We will denote by $(\lieB, i, L)$ such a Cartan operation on $(\Omega^\grast(\lieA, \varE), \hd_\phi)$. A Cartan operation of  a Lie algebra can also be defined in the same way.

Given a Cartan operation, one can define horizontal, invariant and basic elements in $\Omega^\grast(\lieA, \varE)$: $\Omega^\grast(\lieA, \varE)_\hor$ is the graded subspace of horizontal elements (kernel of all the $i_\kX$, for $\kX \in \lieB$), $\Omega^\grast(\lieA, \varE)_\inv$ is the graded subspace of invariant elements (kernel of all the $L_\kX$, for $\kX \in \lieB$), and $\Omega^\grast(\lieA, \varE)_\bas = \Omega^\grast(\lieA, \varE)_\hor \cap \Omega^\grast(\lieA, \varE)_\inv$ is the graded subspace of basic elements.

The kernel $\lieL$ of $\lieA$ defines a natural Cartan operation when the map $i$ is the restriction to $\iota(\lieL)$ of the ordinary inner operation on forms.

\smallskip
Let $\lieA = \tla(\varM, \kg)$ be a trivial Lie algebroid. Then the graded commutative differential algebra $(\Omega^\grast(\lieA), \hd_\lieA)$ is the total complex of the bigraded commutative algebra $\Omega^\grast(\varM) \otimes \exter^\grast \kg^\ast$ equipped with the two differential operators
\begin{align*}
\dd : \Omega^\grast(\varM) \otimes \exter^\grast \kg^\ast &\rightarrow \Omega^{\grast+1}(\varM) \otimes \exter^\grast \kg^\ast,
&
\ds : \Omega^\grast(\varM) \otimes \exter^\grast \kg^\ast &\rightarrow \Omega^\grast(\varM) \otimes \exter^{\grast+1} \kg^\ast,
\end{align*}
where $\dd$ is the de~Rham differential on $\Omega^\grast(\varM)$, and $\ds$ is the Chevalley-Eilenberg differential on $\exter^\grast \kg^\ast$, so that $\hd_\lieA = \dd + \ds$. In the same way, the graded differential Lie algebra $(\Omega^\grast(\lieA, \lieL), \hd)$ is the total complex of the bigraded Lie algebra $\Omega^\grast(\varM) \otimes \exter^\grast \kg^\ast \otimes \kg$ equipped with the differential $\dd$ and the Chevalley-Eilenberg  differential $\ds'$ on $\exter^\grast \kg^\ast \otimes \kg$ for the adjoint representation of $\kg$ on itself, so that $\hd = \dd + \ds'$. We will use the compact notation $(\Omega^\grast_\tla(\varM,\kg), \hd_\tla)$ for this graded differential Lie algebra.

Let $\lieA$ be the Atiyah Lie algebroid of a $G$-principal fiber bundle $\varP$ over $\varM$, and denote by $(\Omega^\grast_\lie(\varP, \kg), \hd)$ its associated differential calculus of forms with values in its kernel. Let $\kg_\equ = \{ \xi^\varP \oplus \xi \ / \ \xi \in \kg \} \subset \Gamma(T\varP \oplus (\varP \times \kg))$: it is a Lie sub algebra of the trivial Lie algebroid $\tla(\varP, \kg)$, and, as such, it induces a natural Cartan operation on the differential complex $(\Omega^\grast_\tla(\varP,\kg), \hd_\tla)$. Let us denote by $(\Omega^\grast_\tla(\varP,\kg)_{\kg_\equ}, \hd_\tla)$ the differential graded subcomplex of basic elements. 

It has been proved in \cite{LazzMass12a} that when $G$ is connected and simply connected, $(\Omega^\grast_\lie(\varP, \kg), \hd)$ and $(\Omega^\grast_\tla(\varP,\kg)_{\kg_\equ}, \hd_\tla)$ are isomorphic as differential graded complexes. This describes the differential calculus of an Atiyah Lie algebroid as the subspace of basic forms in $\Omega^\grast(\varP) \otimes \exter^\grast \kg^\ast \otimes \kg$. 

This description must be compared with the description of sections of an associated fiber bundle as equivariant maps on the principal fiber bundle with valued in the space which is the fiber model of the associated fiber bundle.

\subsection{Gauge field theories}
\label{sec Gauge field theories}

Let us consider the theory of connections in this framework. There is first an ordinary notion of connection \cite{Mack05a}, defined on a  transitive Lie algebroid $\lieA \xrightarrow{\rho} \Gamma(T\varM)$, as a splitting $\nabla : \Gamma(T\varM) \rightarrow \lieA$ of the short exact sequence \eqref{eq-sectransitiveliealgebroid} as $C^\infty(\varM)$-modules. Its curvature is defined to be the obstruction to be a morphism of Lie algebras: $R(X,Y) = [\nabla_X, \nabla_Y] - \nabla_{[X,Y]}$. 

To such a connection there is an associated $1$-form defined as follows: for any $\kX \in \lieA$, let $X = \rho(\kX)$, then $\kX - \nabla_X \in \ker \rho$, so that there is an element $\alpha(\kX) \in \lieL$ such that
\begin{equation}
\label{eq-decompositionwithconnection}
\kX = \nabla_X - \iota \circ \alpha(\kX).
\end{equation}
The map $\alpha : \lieA \rightarrow \lieL$ is a morphism of $C^\infty(\varM)$-modules, so that $\alpha \in \Omega^1(\lieA, \lieL)$. It is normalized on $\iota \circ \lieL$ by the relation $\alpha \circ \iota(\ell) = -\ell$ for any $\ell \in \lieL$. Conversely, any $1$-form $\alpha \in \Omega^1(\lieA, \lieL)$ normalized as before defines a connection on $\lieA$. The $2$-form $\hR = \hd \alpha + \frac{1}{2} [\alpha, \alpha]$ is horizontal for the Cartan operation of $\lieL$ on $(\Omega^\grast(\lieA, \lieL), \hd)$, and with obvious notations, one has $\iota\circ \hR(\kX, \kY) = \iota \left( (\hd \alpha)(\kX, \kY) + [\alpha(\kX), \alpha(\kY)]\right) = R(X,Y)$. $\hR \in \Omega^2(\lieA, \lieL)$ is called the curvature $2$-form of $\nabla$. It satisfies the Bianchi identity $\hd \hR + [\alpha, \hR] = 0$.

For the transitive Lie algebroid $\kD(\varE)$ of derivations of a vector bundle, a connection $\nabla^\varE$ associates to any $X \in \Gamma(T\varM)$ a map $\nabla^\varE_X : \Gamma(\varE) \to \Gamma(\varE)$, and all the relations of  \eqref{eq-definition-rules-nabla} are satisfied. This is then an ordinary covariant derivative on $\varE$. 

More generally, let $\phi : \lieA \to \kD(\varE)$ be a representation of $\lieA$ and $\alpha$ the connection $1$-form of a connection $\nabla$ on $\lieA$. Then for any $\psi \in \Gamma(\varE)$, $\lieA \ni \kX \mapsto \phi(\kX)\cdotaction \psi + \phi_\lieL(\alpha(\kX)) \psi$ vanishes for $\kX = \iota(\ell)$ for any $\ell \in \lieL$, so that $\nabla^\varE_X \psi = \phi(\kX)\cdotaction \psi + \phi_\lieL(\alpha(\kX)) \psi$ is well-defined with $X = \rho(\kX)$, and it is a covariant derivative on $\varE$, in the sense of \eqref{eq-definition-rules-nabla}.

A connection on the Atiyah Lie algebroid of a principal fiber bundle $\varP$ associates to $X \in \Gamma(T\varM)$ a right invariant vector field $\nabla_X \in \Gamma_G(T\varP)$. This corresponds to the usual horizontal lift $X \mapsto X^h$ defined by a connection $\omega$ on $\varP$. Suppose now that $G$ is connected and simply connected. The connection $1$-form $\omega$ on $\varP$ is an element of $\Omega^1(\varP) \otimes \kg$, and so of $\Omega^1(\varP) \otimes \exter^0 \kg^\ast \otimes \kg$, which satisfies \eqref{eq-relations-def-omega}. Let $\theta \in \exter^1 \kg^\ast \otimes \kg$ be the Maurer-Cartan $1$-form on $G$, which can be considered as an element in $C^\infty(\varP) \otimes \exter^1 \kg^\ast \otimes \kg$. The difference $\omega - \theta$ then belongs to $\Omega^1_\tla(\varP,\kg)$, and using the properties of $\omega$ and $\theta$, it is easy to show that it is $\kg_\equ$-basic. As a basic element in $\Omega^1_\tla(\varP,\kg)$, it identifies with $\alpha \in \Omega^1_\lie(\varP, \kg)$ in the correspondence described before (see \cite{LazzMass12a}  for details).

As a global object on $\varM$, the generalized $1$-form $\alpha$ associated to the connection $1$-form $\omega$ on $\varP$ completes the last description proposed in \ref{sec Ehresmann connections} on connections and curvatures in ordinary differential geometry. Using local descriptions of the Atiyah transitive Lie algebroid $\Gamma_G(T\varP)$ (in terms of local trivializations of $\varP$), the local descriptions of $\alpha$ are given by $A_i - \theta$, where the $A_i$'s are the local trivialization $1$-forms of $\omega$.

The three above examples show that this ordinary notion of connections on transitive Lie algebroid is close to the geometric notion of connections described in \ref{sec Ehresmann connections}.

This notion of connections on Lie algebroids admits generalizations under different names (\cite{Fern02a} and references therein): $\lieA$-connections or $\lieA$-derivatives. Here we introduce a definition proposed in \cite{LazzMass12a}, which is more restrictive than the other ones, but which fits perfectly with the ambition to promote a transitive Lie algebroid to an infinitesimal version of a principal fiber bundle. Such a principal fiber bundle supports the primary notion of connection and defines completely the gauge group, and all these notions are transferred to associated vector bundle (``representations''). We will do the same for transitive Lie algebroids.

A generalized connection $1$-form on the transitive Lie algebroid $\lieA$ is then a $1$-form $\homega \in \Omega^1(\lieA, \lieL)$, and its curvature is the $2$-form $\hR = \hd \homega + \frac{1}{2}[\homega, \homega] \in \Omega^2(\lieA, \lieL)$.

Since $\lieA$ is a kind of infinitesimal version of a principal fiber bundle, there is no notion of gauge transformations as ``finite'' transformations, but we can identify a Lie algebra of infinitesimal gauge transformations to be $\lieL$. There is at least two motivations for that. First, let $\phi : \lieA \to \kD(\varE)$ be a representation of $\lieA$. Then, for any $\xi \in \lieL$, $\phi_\lieL(\xi)$ defines an infinitesimal gauge transformation on $\varE$. Notice that the gauge group of $\varE$ is well defined as $\Aut(\varE)$, the (vertical) automorphisms of $\varE$. Secondly, for an Atiyah transitive Lie algebroid associated to a principal fiber bundle $\varP$, the kernel $\lieL = \Gamma_G(\varP, \kg)$ identifies as the Lie algebra of infinitesimal gauge transformations on $\varP$. These two examples motivate also the following definition.

The action of an infinitesimal gauge transformation $\xi \in \lieL$ on a generalized connection $1$-form $\homega$ is defined to be the $1$-form $\homega^\xi = \homega + (\hd \xi + [\homega, \xi]) + O(\xi^2)$.

An ordinary connection $\nabla$ on $\lieA$ defines a $1$-form $\alpha$ normalized on $\iota(\lieL)$. This $1$-form then defines a generalized connection on $\lieA$. This implies that the space of ordinary connections on $\lieA$ is contained in the space of generalized connection $1$-forms, and this inclusion is compatible with the notions of curvature and (infinitesimal) gauge transformations.

Let $\phi : \lieA \to \kD(\varE)$ be a representation of $\lieA$, and let $\homega \in \Omega^1(\lieA, \lieL)$ be a generalized connection on $\lieA$. Then its covariant derivative $\hnabla$ on $\varE$ is defined for any $\kX \in \lieA$ as
\begin{equation*}
\hnabla^\varE_\kX = \phi(\kX) + \iota \circ \phi_\lieL \circ \homega(\kX).
\end{equation*}
This is a differential operator on $\Gamma(\varE)$, and one has $[\hnabla^\varE_{\kX}, \hnabla^\varE_{\kY}] - \hnabla^\varE_{[\kX, \kY]} = \phi_\lieL \circ \hR(\kX,\kY)$ for any $\kX, \kY \in \lieA$.

To summarize, we get the following diagram with the structures defined above:
\begin{equation*}
\xymatrix
{
{\algzero} \ar[r] & {\lieL} \ar[r]_-{\iota} \ar[d]_-{\phi_\lieL}  & {\lieA} \ar@/_0.7pc/[l]_-{\homega}  \ar[r]^-{\rho} 
\ar[d]^-{\phi}_-{\hnabla^\varE}
& {\Gamma(T\varM)} \ar[r] \ar@{=}[d]  & {\algzero}
\\
{\algzero} \ar[r] & {\algA(\varE)} \ar[r]^-{\iota} & {\kD(\varE)} \ar[r]^-{\sigma} & {\Gamma(T\varM)} \ar[r] & {\algzero}
}
\end{equation*}
$\hnabla^\varE$ is often called a generalized representation, in the sense that it is not compatible with the Lie brackets.

Generalized connections of Atiyah Lie algebroids can be described as $\kg_\equ$-basic $1$-forms in $\Omega^1_\tla(\varP,\kg)$:
\begin{equation}
\label{eq-split-omegaphi}
\homega =\omega + \phi \in \Omega^1_\tla(\varP,\kg) = (\Omega^1(\varP) \otimes \kg) \oplus (C^\infty(\varP) \otimes \exter^1 \kg^\ast \otimes \kg).
\end{equation}
where $\omega$ and $\phi$ are $\kg_\equ$-invariant, but $\omega$ is not necessarily a connection $1$-form on $\varP$ and $\phi$ is not necessarily related to the Maurer-Cartan form on $G$. Atiyah Lie algebroids admit a notion of (finite) gauge transformations as elements in the ordinary gauge group $\caG(\varP)$ of $\varP$. This shows that the theory of generalized connections on Atiyah Lie algebroids is a close extension of the theory of ordinary connections on $\varP$.

A general theory of metrics, Hodge star operators, and integrations on transitive Lie algebroids has been developed in \cite{FourLazzMass13a}, which permits to write explicit gauge invariant actions for generalized connections and its coupling, via a covariant derivative, to matter fields in a representation $\varE$ of $\lieA$.

A metric on $\lieA$ is a symmetric, $C^\infty(\varM)$-linear map $\hg : \lieA \otimes_{C^\infty(\varM)} \lieA \rightarrow C^\infty(\varM)$. Under certain non degeneracy conditions, such a metric decomposes in a unique way into three pieces:
\begin{enumerate}
\item a metric $g$ on $\varM$,
\item a metric $h$ on the vector bundle $\varL$ such that $\lieL = \Gamma(\varL)$,
\item an ordinary connection $\nabladot$ on $\lieA$, with associated generalized $1$-form $\omegadot$.
\end{enumerate}
A Hodge star operator $\hodgeast$ can then be defined, as well as an integration along the inner part $\lieL$, which, combined with the integration on $\varM$ against the measure $\dvol_g$, produces a global integration $\int_\lieA$. The gauge invariant action is then defined as
\begin{equation}
\label{eq-functionalactiongauge}
\Act_\text{Gauge}[\homega] = \int_\lieA h(\hR,\hodgeast \hR),
\end{equation}
where $\hR$ is the curvature of $\homega$. In order to understand the content of this action functional, one has to introduce the following elements:
\begin{itemize}
\item $\rke = \homega \circ \iota + \Id_{\lieL}$ is an element of $\End(\lieL)$, which contains the degrees of freedom of $\homega$ along $\lieL$, that is the algebraic part of $\homega$;

\item $R_\rke : \lieL \times \lieL \rightarrow \lieL$ is the obstruction for $\rke\in\End(\lieL)$ to be an endomorphism of Lie algebras: $R_\rke(\gamma,\eta) = [ \rke(\gamma) , \rke(\eta) ] - \rke([\gamma,\eta])$ for any $\gamma, \eta \in \lieL$;

\item $\omega = \homega + \rke(\omegadot)$ is the generalized $1$-form of an ordinary connection $\nabla$ on $\lieA$, which contains the degree of freedom of $\homega$ along $\varM$, that is the geometric part of $\homega$;

\item $(\caD_X \rke)(\gamma) = [\nabla_X, \rke(\gamma)] - \rke([\nabladot_X, \gamma])$ is a covariant derivative of $\rke$ along the ordinary connections $\nabla$, for any $X \in \Gamma(T\varM)$ and $\gamma \in \lieL$;

\item $\hF = R - \rke \circ \Rdot \in \Omega^2(\varM, \varL)$, in which $\Rdot, R \in \Omega^2(\varM, \varL)$ are the curvature $2$-forms of the ordinary connections $\omegadot$ and $\omega$.
\end{itemize}
Then the curvature of $\homega$ decomposes into three parts: $\hR = \rho^\ast \hF - (\rho^\ast \caD \rke)\circ \omegadot + \omegadot^\ast R_\rke$, and $\Act_\text{Gauge}[\homega]$ is a sum of the squares of these three pieces.

When $\homega$ is an ordinary connection (normalized on $\lieL$), $\rke = 0$, so that $R_\rke = 0$, $\omega = \homega$, $\caD_X \rke=0$, $\hF = R$. On an Atiyah Lie algebroid, the action functional then reduces exactly to the Yang-Mills action \eqref{eq-action-YM}. 

The gauge theories obtained in this way are of Yang-Mills-Higgs type: the fields in the ordinary connection $\omega$ are Yang-Mills-like fields, and the $\rke$'s fields behave as scalar fields which exhibit a SSBM. Indeed, the potential for these fields is the square of $\omegadot^\ast R_\rke$, and it vanishes when $\rke$ is a Lie algebra morphism. This can occur for instance when $\rke = \Id_\lieL$, and this non zero configuration, once reported into the square of the covariant derivative $(\rho^\ast \caD \rke)\circ \omegadot$, induces mass terms for the (geometric) fields contained in $\omega$. There is a similar decomposition of the action functional associated to the minimal coupling with matter fields, and the algebraic part $\rke$ of $\homega$ induces also mass terms for these matter fields.

\section{Conclusion}

It is worthwhile to notice similarities between some of the constructions presented at the end of section \ref{sec Derivation-based NCG} on the endomorphism algebra of a $SU(n)$-vector bundle, and some of the constructions presented in section \ref{sec Gauge field theories} on transitive Lie algebroids. In particular, they share the following structures:
\begin{itemize}
\item both constructions make apparent a short exact sequence of Lie algebras and $C^\infty(\varM)$-modules, \eqref{eq-ses-endo} and \eqref{eq-sectransitiveliealgebroid};

\item the notion of ordinary connections corresponds in both situations to a splitting of these short exact sequences;

\item the connection $1$-form associated to such an ordinary connection uses in both situations the defining relation \eqref{eq-decompositionwithconnection};

\item gauge field theories written in both situations are of the Yang-Mills-Higgs type, and they remain close to ordinary gauge field theories in their formulation.
\end{itemize}
These similarities are not pure coincidence. They reflect a result proved in \cite{LazzMass12a}, where we use the fact that \eqref{eq-ses-endo} defines $\der(\algA)$ as a transitive Lie algebroid. The following three spaces are isomorphic:
\begin{enumerate}
\item The space of generalized connection $1$-forms on the  transitive Lie algebroid $\der(\algA)$.
\item The space of generalized connection $1$-forms on the Atiyah Lie algebroid $\Gamma_G(T\varP)$, where $\varP$ is the principal fiber bundle underlying the geometry of the endomorphism algebra $\algA$.
\item The space of traceless noncommutative connections on the right $\algA$-module $\modM = \algA$
\end{enumerate}
The isomorphisms are compatible with curvatures and (finite) gauge transformations. Moreover, these spaces contain the ordinary connections on $\varP$, and the inclusion is compatible with curvatures and gauge transformations. 

This result shows that the two generalizations of the ordinary notion of connections proposed in \ref{sec Derivation-based NCG} and \ref{sec Gauge field theories} are more or less the same, and they extend in the same ``direction'' the usual notion of connection introduced in the geometrical framework described in \ref{sec Ehresmann connections}. In both constructions, the generalized connections split into two parts, see \eqref{eq-split-ab} and \eqref{eq-split-omegaphi}: a Yang-Mills type vector field $\ka$ or $\omega$, and some scalar fields $\kb$ or $\phi$. The corresponding Lagrangians \eqref{eq-ncg-act} and \eqref{eq-functionalactiongauge} provide for free a (purely algebraic) quadratic potential for the scalar fields  which allows a SSBM with mass generation. This cures the mathematical weakness of the SM stressed in the introduction.

As explained before, \nc geometry restricts the possible gauge group to the automorphism group of the algebra $\algA$, but using Atiyah Lie algebroids, this restriction is no more true, since any principal fiber bundle can be considered.

More generally, the approaches described in subsections~\ref{sec Ehresmann connections}, \ref{sec Spectral triples}, \ref{sec Derivation-based NCG} and \ref{sec Gauge field theories} share a common structure which  appears under the form of ``sequences'' such as \eqref{eq-principal-bundle}, \eqref{eq-ses-aut-LM}, \eqref{eq-ses-Cartan}, \eqref{eq-ses-automorphism-groups}, \eqref{eq-sesderivations}, \eqref{eq-sectransitiveliealgebroid} and \eqref{eq-ses-atiyah}. All of them reproduce the same following pattern:


\begin{equation}
\label{eq-universal-sequence}
\begin{tikzpicture}[node distance=3.5ex and 5em, auto, baseline=(global.west)]
\node[conceptbox] (alg) {\centeredlines{Algebraic\\ structure}};
\node[conceptbox,right= of alg] (global) {\centeredlines{Global\\ structure}};
\node[conceptbox,right= of global] (geom) {\centeredlines{Geometric\\ structure}};

\path[flechestructure] (alg) -- node[label] {inclusion} (global);
\path[flechestructure] (global) -- node[label] {projection} (geom);
\end{tikzpicture} 
%
\end{equation}
This pattern embodies the characterization of gauge field theories exposed in the introduction. The ``geometric structure'' in this diagram represents the basic symmetries induced by the base (space-time) manifold $\varM$ (diffeomorphisms, change of coordinate systems), while the ``algebraic structure'' is a supplementary ingredient on top of $\varM$ (a group, a Lie algebra…) from which emerges the characterization of gauge fields in the theory (mainly through representation theory). The ``global structure'' in the middle encodes all the symmetries of the theory, under a structure which can not be splited in general (group of all the automorphisms of a principal fiber bundle, automorphisms of an associative algebra, transitive Lie algebroid $\lieA$…). The local dependance of gauge transformations in a gauge field theory is then the result of the (geometric) implementation of an algebraic structure on top of a base manifold.

Einstein's theory of gravitation can be written in terms of purely geometric structures, on the right of the diagram. Using a suitable formalism, for instance reductive Cartan geometries, this construction can be \emph{lifted} to a ``global structure'', in a theory which contains some extra degrees of freedom in new fields, submitted to a (new) gauge symmetry with the same amount of degrees of freedom. On the contrary, Yang-Mills(-Higgs) type theories are defined using the algebraic structure on the left and its associated symmetries (vertical automorphisms of a principal fiber bundle, inner symmetries in \nc geometry, kernel of a transitive Lie algebroid…).

This pattern is at the core of gauge field theories, and reflects the most general mathematical structure underlying these theories, at least at the classical level. The relevance of this pattern to quantize gauge field theories is another important issue which deserves a more special attention.

\bibliography{biblio-Copernicus-center}

\end{document}